# Color, Composition, and Thermal Environment of Kuiper Belt Object (486958) Arrokoth

W.M. Grundy[1,2*], M.K. Bird[3,4], D.T. Britt[5], J.C. Cook[6], D.P. Cruikshank[7], C.J.A. Howett[8], S. Krijt[9], I.R. Linscott[10], C.B. Olkin[8], A.H. Parker[8], S. Protopapa[8], M. Ruaud[7], O.M. Umurhan[7,11], L.A. Young[8], C.M. Dalle Ore[7,11], J.J. Kavelaars[12,13], J.T. Keane[14], Y.J. Pendleton[7], S.B. Porter[8], F. Scipioni[7,11], J.R. Spencer[8], S.A. Stern[8], A.J. Verbiscer[15], H.A. Weaver[16], R.P. Binzel[17], M.W. Buie[8], B.J. Buratti[18], A. Cheng[16], A.M. Earle[17], H.A. Elliott[19], L. Gabasova[20], G.R. Gladstone[19], M.E. Hill[16], M. Horanyi[21], D.E. Jennings[22], A.W. Lunsford[22], D.J. McComas[23], W.B. McKinnon[24], R.L. McNutt Jr.[16], J.M. Moore[7], J.W. Parker[8], E. Quirico[20], D.C. Reuter[22], P.M. Schenk[25], B. Schmitt[20], M.R. Showalter[11], K.N. Singer[8], G.E. Weigle II[26], A.M. Zangari[8]

1. Lowell Observatory, Flagstaff AZ 86001 USA
2. Department of Astronomy and Planetary Science, Northern Arizona University, Flagstaff AZ 86011 USA
3. Argelander-Institut für Astronomie, University of Bonn, Bonn D-53121, Germany
4. Rheinisches Institut für Umweltforschung, Universität zu Köln Cologne 50931, Germany
5. University of Central Florida, Orlando FL 32816 USA
6. Pinhead Institute, Telluride CO 81435 USA
7. NASA Ames Research Center, Moffett Field CA 94035 USA
8. Southwest Research Institute, Boulder CO 80302 USA
9. Hubble Fellow, Steward Observatory, University of Arizona, Tucson AZ 85719 USA
10. Stanford University, Stanford CA 94305 USA
11. Carl Sagan Center, Search for Extraterrestrial Intelligence Institute, Mountain View CA 94043 USA
12. National Research Council, Victoria BC V9E 2E7 Canada
13. Department of Physics and Astronomy, University of Victoria, Victoria BC V8W 2Y2 Canada
14. California Institute of Technology, Pasadena CA 91125 USA
15. University of Virginia, Charlottesville VA 22904 USA
16. Johns Hopkins University Applied Physics Laboratory, Laurel MD 20723 USA
17. Massachusetts Institute of Technology, Cambridge MA 02139 USA
18. NASA Jet Propulsion Laboratory, La Cañada Flintridge CA 91011 USA
19. Southwest Research Institute, San Antonio TX 78238 USA
20. Institut de Planétologie et d'Astrophysique de Grenoble, Centre National de la Recherche Scientifique, Université Grenoble Alpes, Grenoble France
21. University of Colorado, Boulder CO 80309 USA
22. NASA Goddard Space Flight Center, Greenbelt MD 20771 USA
23. Princeton University, Princeton NJ 08544 USA

24. Washington University of St. Louis, St. Louis MO 63130 USA
25. Lunar and Planetary Institute, Houston TX 77058 USA
26. Big Head Endian LLC, Leawood KS 67019 USA
*Correspondence to: w.grundy@lowell.edu

**The outer Solar System object (486958) Arrokoth (provisional designation 2014 MU$_{69}$) has been largely undisturbed since its formation.  We study its surface composition using data collected by the New Horizons spacecraft.  Methanol ice is present along with organic material, which may have formed through radiation of simple molecules.  Water ice was not detected.  This composition indicates hydrogenation of carbon monoxide-rich ice and/or energetic processing of methane condensed on water ice grains in the cold, outer edge of the early Solar System.  There are only small regional variations in color and spectra across the surface, suggesting Arrokoth formed from a homogeneous or well-mixed reservoir of solids.  Microwave thermal emission from the winter night side is consistent with a mean brightness temperature of 29 ± 5 K.**

The New Horizons spacecraft flew past (486958) Arrokoth at the beginning of 2019 *(1)*.  Arrokoth rotates with a 15.9 hour period about a spin axis inclined 99.3° to the pole of its 298 year orbit at a mean distance from the Sun of 44.2 AU *(2, 3)*.  Its near-circular orbit, with a mean eccentricity of 0.03 and inclination of 2.4° to the plane of the Solar System, makes it a Kuiper belt object (KBO) and more specifically, a member of the "kernel" sub-population of the cold classical KBOs (CCKBOs) *(4)*.  CCKBOs have distinct origins and properties from KBOs on more excited orbits, which are thought to have formed closer to the Sun before being perturbed outward by migrating giant planets early in Solar System history *(5)*.  CCKBOs still orbit where they formed in the protoplanetary nebula, the accretion disk of gas and dust around the young Sun.  They have a high fraction of binary objects *(6)*, a uniformly red color distribution *(7, 8)*, a size-frequency distribution deficient of large objects *(9, 10)*, and higher albedos *(11, 12)*.  These properties arise from the environment at the outermost edge of the protoplanetary nebula, from a distinct history of subsequent evolution of CCKBOs compared to other KBOs, or of some combination of these two.  Arrokoth provides a record of the process of forming planetesimals, the first generation of gravitationally bound bodies, that has been minimally altered by subsequent processes such as heating and impactor bombardment *(3)*.  Its distinctive bi-lobed, 35 km-long shape with few impact craters favors formation via rapid gravitational collapse, rather than scenarios involving more gradual accretion via piece-wise agglomeration of dust particles to assemble incrementally larger aggregates *(13)*.  We study Arrokoth's color, composition, and thermal environment using data from the New Horizons flyby, and discuss the resulting implications for its formation and subsequent evolution.

## Instruments and Data

New Horizons encountered Arrokoth when it was 43.28 AU from the Sun, collecting data with its suite of scientific instruments. Color and compositional remote sensing data were provided by the Ralph color camera and imaging spectrometer, sensitive to wavelengths between 0.4 and 2.5 µm *(14)*. Over this wavelength range, all light observed from Arrokoth is reflected sunlight, with the wavelength-dependence of the reflectance indicative of surface composition and texture. Ralph's two focal planes share a single 75 mm aperture telescope using a dichroic beamsplitter. The Multi-spectral Visible Imaging Camera (MVIC) provides panchromatic and color imaging in four color filters, "BLUE" (400-550 nm), "RED" (540-700 nm), "NIR" (780-975 nm), and "CH4" (860-910 nm) *(15)*. The highest spatial resolution MVIC color observation of Arrokoth designated as CA05, was obtained on 2019 January 1 at 5:14 UTC (coordinated universal time), from a range of 17,200 km, at an image scale of 340 meters per pixel, and phase angle 15.5°. This provides more spatial detail than the 860 meters per pixel CA02 MVIC color scan *(1)*.

Ralph's Linear Etalon Infrared Spectral Array (LEISA) images its target scene through a linear variable filter covering wavelengths from 1.2 to 2.5 µm at a spectral resolving power of about 240. To capture each location at each wavelength of the filter, frames are recorded while the spacecraft scans LEISA's field of view across the scene. The highest spatial resolution LEISA observation, designated as CA04, was executed around 4:58 UTC, shortly before the CA05 MVIC observation, from a phase angle 12.6° and mean range of 31,000 km, resulting in a mean image scale of 1.9 km per pixel *(15)*.

New Horizons' panchromatic LOng-Range Reconnaissance Imager [*(16)* LORRI] is co-aligned with Ralph and can record images while the spacecraft is scanning for a Ralph observation. Such LORRI observations, referred to as "riders", are limited to short integration times to minimize image smear from scan motion, but multiple images can be recorded and combined in post-processing, providing for longer effective integration times *(3)*. LORRI rider observations were obtained during both the CA04 and CA05 observations, providing higher spatial resolution context images for the Ralph observations.

New Horizons' Radio Science Experiment [*(17)* REX], was used to observe thermal emission in the X-band (4.2 cm wavelength, 7.2 GHz) from Arrokoth's Sun-oriented face on approach and then from its anti-Sun oriented face on departure. The two REX observations, designated as CA03 and CA08, respectively, were obtained January 1 at mean times of 4:34 and 5:52 UTC, phase angles of 11.9° and 162.0°, and ranges of 52,000 and 16,700 km. At those distances, Arrokoth was unresolved, appearing much smaller than the 1.2° width (at 3 dB) of the high-gain antenna beam. Two independent receivers recorded the radio flux density in different polarizations at a sampling rate of 10 Hz. The REX A receiver recorded right circularly polarized flux while REX B recorded left circularly polarized flux.

## Visible Wavelength Color

The CA02 MVIC color scan *(1)* had shown Arrokoth to be red, but revealed little spatial variation in color. The higher resolution CA05 observation, allows us to better quantify Arrokoth's regional color variations. Fig. 1 shows the CA05 color image, compared with the contemporaneous LORRI panchromatic rider image. Color slopes, computed by fitting a linear model to the MVIC BLUE, RED, and NIR filter reflectance data are shown in Fig. 1C. All of Arrokoth's surface is red in color, with a mean color slope of 27% rise per 100 nm (at 550 nm). This quantification of color slopes is commonly used for KBOs, being convenient for comparison of colors obtained using different filter sets *(18)*. Even in the higher spatial resolution of the CA05 observation, the color distribution is largely uniform across the observed face of Arrokoth, with a standard deviation in slope of only ±2.7% per 100 nm.

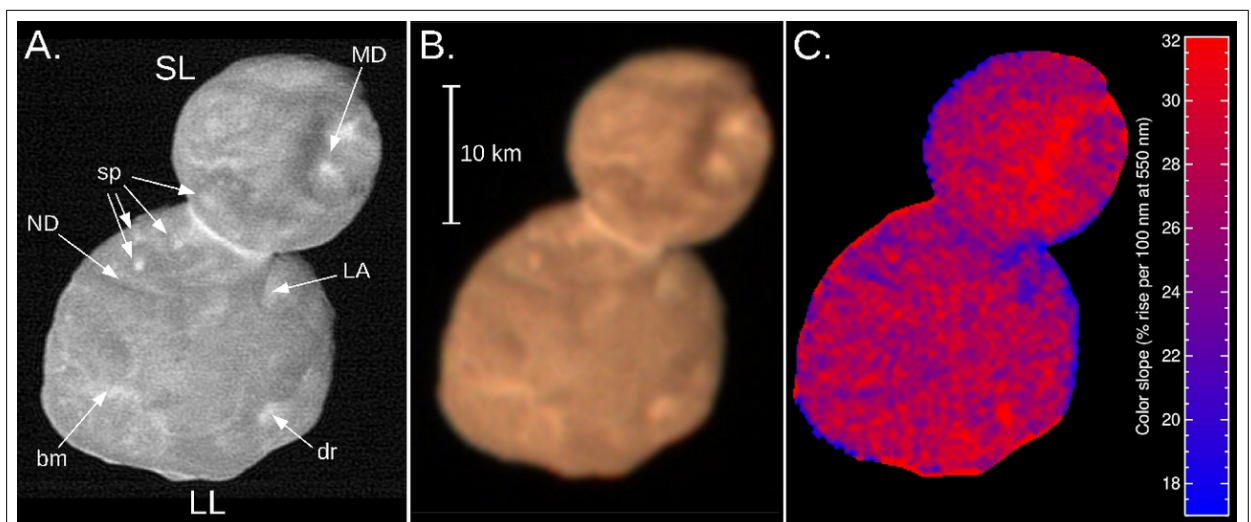

**Figure 1. The CA05 color observation of (486958) Arrokoth.** A. LORRI panchromatic context image obtained as a rider during the CA05 observation, so the geometry is nearly identical to the MVIC observation, but with a finer spatial scale of 83 m pixel$^{-1}$. Abbreviations and informal feature nicknames mentioned in the text are (clockwise from the top): SL smaller lobe, MD Maryland, LA Louisiana, dr depressed region, LL larger lobe, bm bright material, ND North Dakota, and sp spots. These features can be seen more clearly in higher resolution LORRI images [*(3)*, their figure 1]. B. Color observation CA05, at a spatial scale of 340 m pixel$^{-1}$. The BLUE filter (400-550 nm) is displayed in blue, RED filter (540-700 nm) in green, and NIR filter (780-975 nm) in red. C. Color slope map obtained by fitting a linear model to the BLUE, RED, and NIR reflectance values.

Subtle regional color differences correspond to specific geological and albedo features discussed in a companion paper *(3)*. The smaller lobe (SL) appears slightly redder on average than the larger lobe (LL), 28±2% average slope versus 27±2% for LL, where the ±2 values represent the variance across each lobe, rather than the uncertainty in the measurement of the mean slopes, which is much smaller for averages over many pixels *(15)*. That difference appears to be mostly due to the redder rim (color slope 30±2%) of a 6 km diameter depression on SL, a possible impact crater informally identified as Maryland (MD, all place names are informal nicknames). Statistically significant color differences tend to be that small or smaller in spatial scale. Several slightly less red regions appear as blue in the color scale used in Fig. 1C. These include the brighter neck region where the two lobes intersect (25±1% slope), and several regions on LL. Two that were not resolved in the earlier color data are a depression near the neck nicknamed Louisiana (LA, 23±2% slope) and a linear depression or groove North Dakota (ND, 24±1% slope), labeled in Fig. 1A. Bright material (bm) in the geological map *(3)* is in some places more and in others less red than average, suggesting that unit is composed of two or more distinct materials. Another depressed region (dr) on LL is slightly redder than the average (29±2% slope). Some bright spots (sp) appear to have distinct colors as well, though they are not all the same. Some are a little redder than average while others are a little less red. The lack of a consistent color pattern for these spots suggests they may have resulted from delivery of diverse material in impacts rather than by impact excavation of a uniform subsurface material. However, the nature of these spots remains ambiguous *(3)*.

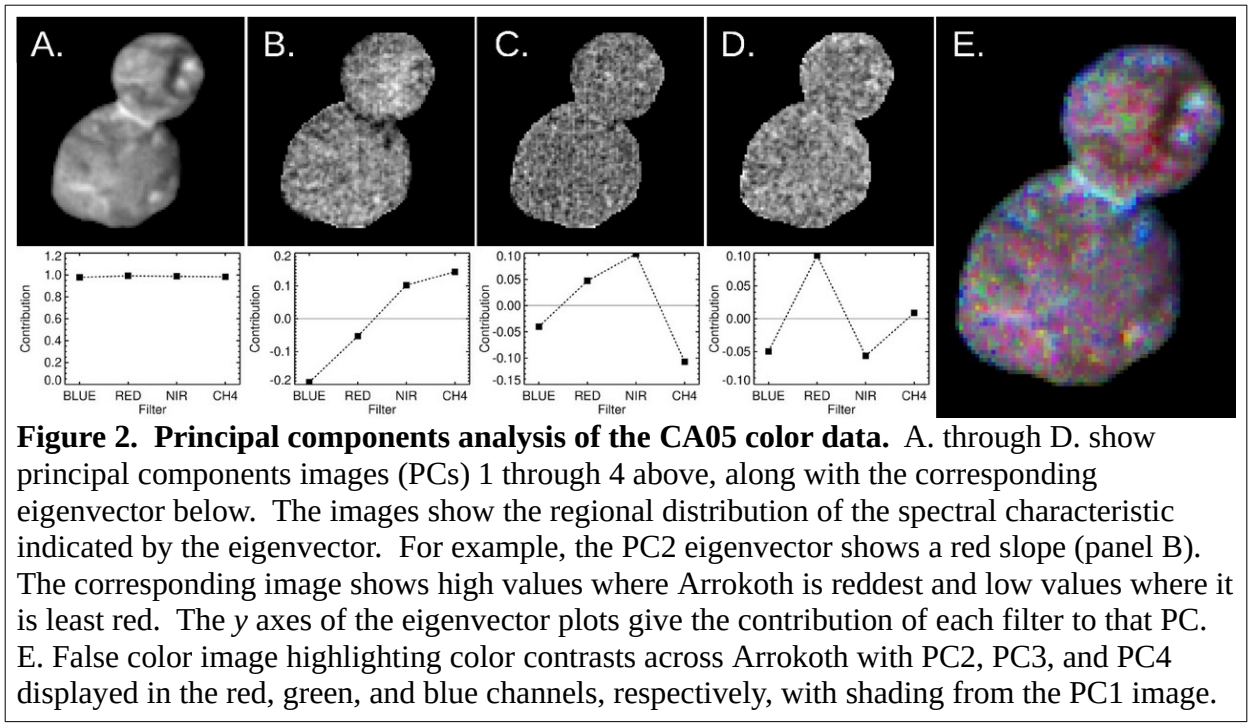

**Figure 2. Principal components analysis of the CA05 color data.** A. through D. show principal components images (PCs) 1 through 4 above, along with the corresponding eigenvector below. The images show the regional distribution of the spectral characteristic indicated by the eigenvector. For example, the PC2 eigenvector shows a red slope (panel B). The corresponding image shows high values where Arrokoth is reddest and low values where it is least red. The $y$ axes of the eigenvector plots give the contribution of each filter to that PC. E. False color image highlighting color contrasts across Arrokoth with PC2, PC3, and PC4 displayed in the red, green, and blue channels, respectively, with shading from the PC1 image.

We performed a principal components analysis (PCA) of the color data (Fig. 2). This analysis projects the data into an orthogonal basis set, with the first axis corresponding to the axis of maximum variance within the data. The second axis corresponds to the maximum remaining variance after collapsing the data along the first axis, and so forth. Principal component 1 (PC1, Fig. 2A) captures variations in brightness due to lighting and albedo, accounting for 97% of the total variance in the data. The corresponding eigenvector is flat across all filters. PC2 corresponds to redness (Fig. 2B), closely resembling the color slope map in Fig. 1C, and accounts for 1.8% of the total variance. PC3 and PC4 correspond to contrasts between the NIR and CH4 filters and to spectral curvature between BLUE, RED, and NIR filters, respectively. They account for only 1% of the variance between them, much of it due to image noise rather than real color variations across the surface of Arrokoth.

Red coloration on planetary bodies is often attributed to the presence of tholins *(19)*. These are a broad class of refractory macromolecular polymer-like organic solids, commonly produced in laboratory simulations of energetic radiation acting on various combinations of simpler molecules *(20-22)*. The precursors can be in gaseous form *(23)* or frozen solid *(24, 25)*. Fig. 3 compares the color of Arrokoth with other KBOs and related populations. Arrokoth's color slope is consistent with other CCKBOs *(8)*. It is less red than the reddest KBOs and the red group of Centaurs *(26)*, where the red coloration is generally interpreted as due to tholins *(19)*. Arrokoth is much more red than the gray Centaurs and various classes of asteroids *(27)*, including the red D-type asteroids that dominate the Jupiter Trojan population *(28)*. Using the Sloan *g, r,* and *z* filters, KBO colors can be placed on a color-color plot of *g–r* and *r–z* color differences, which has revealed a distinction between CCKBOs and more dynamically excited KBO populations *(8)*. The latter appear to follow two color tracks (*(29)* Fig. 3B) while the CCKBOs cluster below and to the right, owing to their red slopes becoming less steep at *z*-band wavelengths (0.82 to 0.96 µm). Arrokoth's red slope continues into MVIC's NIR band (0.78 to 1 µm). After converting to Sloan colors, Arrokoth lies above the main clump of CCKBO colors in Fig. 3B, although this is consistent with the overall CCKBO color distribution.

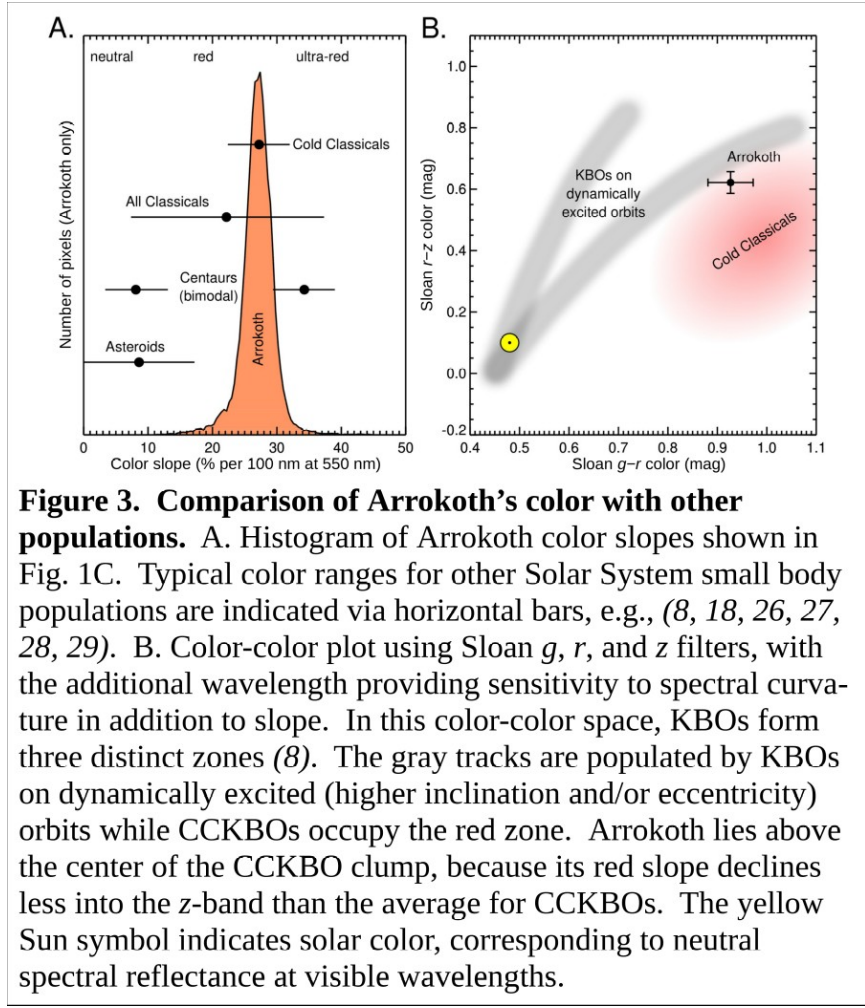

**Figure 3. Comparison of Arrokoth's color with other populations.** A. Histogram of Arrokoth color slopes shown in Fig. 1C. Typical color ranges for other Solar System small body populations are indicated via horizontal bars, e.g., *(8, 18, 26, 27, 28, 29)*. B. Color-color plot using Sloan *g, r,* and *z* filters, with the additional wavelength providing sensitivity to spectral curvature in addition to slope. In this color-color space, KBOs form three distinct zones *(8)*. The gray tracks are populated by KBOs on dynamically excited (higher inclination and/or eccentricity) orbits while CCKBOs occupy the red zone. Arrokoth lies above the center of the CCKBO clump, because its red slope declines less into the *z*-band than the average for CCKBOs. The yellow Sun symbol indicates solar color, corresponding to neutral spectral reflectance at visible wavelengths.

# Near-Infrared Spectral Reflectance

The LEISA data were processed into a spectral cube, with spatial dimensions along two axes and wavelength along the third axis *(15)*. The cube-building algorithm we adopted accounts for changes in spacecraft range over the course of the CA04 LEISA scan [unlike *(1)*]. The spatial resolution of the LEISA data is considerably coarser than the corresponding LORRI rider (Fig. 4), and the signal/noise ratio is considerably lower. Noise in the LEISA data is dominated by instrumental effects, which are as large as the signal from sunlit areas on Arrokoth, so spectral features can only be revealed by averaging over multiple pixels and/or wavelengths. The spatially averaged spectrum (Fig. 4C) lacks the strong absorption features that were seen in the Pluto system, e.g., *(30)*. The red color slope seen in the visible flattens with increasing wavelength to become neutral by around 1.5 μm. We constructed Hapke reflectance models *(15, 31)* for various combinations of potential surface components. The data support inclusion of amorphous carbon and tholins in the models. Combinations of these materials match the overall albedo and spectral shape. Although we used tholins made in conditions simulating the atmosphere of the moon Titan *(20)*, the data do not support singling out any of the various tholins with published optical constants. None have been made under simulated outer nebular conditions, so they may not be particularly good analogs for the tholins on Arrokoth. Amorphous carbon has no diagnostic spectral features in this wavelength range, so it cannot be specifically identified. Any dark, spectrally neutral material would be equally consistent with the data.

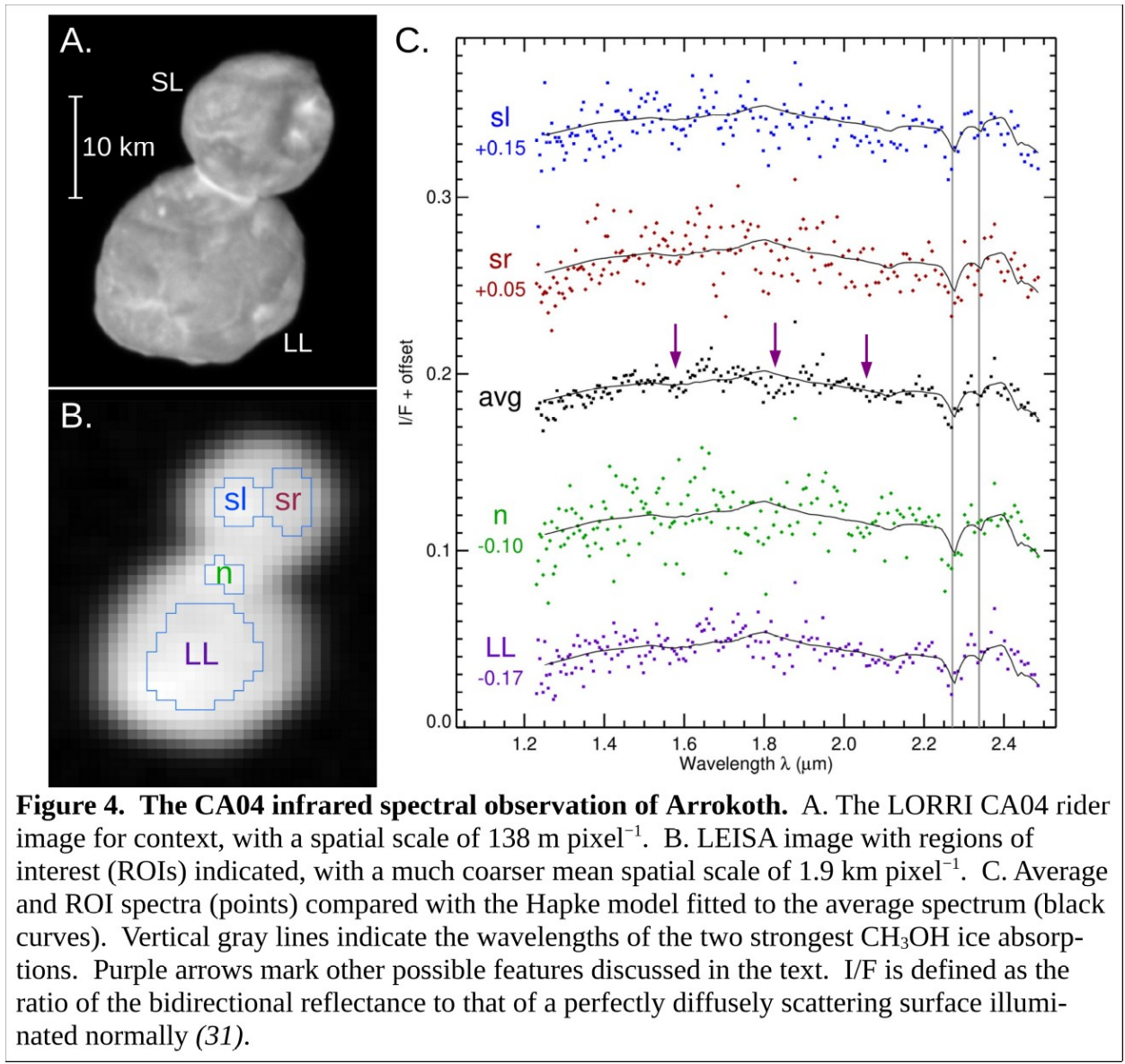

**Figure 4. The CA04 infrared spectral observation of Arrokoth.** A. The LORRI CA04 rider image for context, with a spatial scale of 138 m pixel$^{-1}$. B. LEISA image with regions of interest (ROIs) indicated, with a much coarser mean spatial scale of 1.9 km pixel$^{-1}$. C. Average and ROI spectra (points) compared with the Hapke model fitted to the average spectrum (black curves). Vertical gray lines indicate the wavelengths of the two strongest $CH_3OH$ ice absorptions. Purple arrows mark other possible features discussed in the text. I/F is defined as the ratio of the bidirectional reflectance to that of a perfectly diffusely scattering surface illuminated normally *(31)*.

Shallow absorption bands can be discerned near 1.5–1.6, 1.8, 2.0–2.1, 2.27 and 2.34 µm (Fig. 4C). Adding ices of methanol ($CH_3OH$), water ($H_2O$), and ammonia ($NH_3$) to the models enables them to match many of these features, but not the one at 1.8 µm. With the low signal to noise ratio of the LEISA spectrum, even in the global average, we must consider the information content and limit the model free parameters to those that can be statistically justified. Of the molecular ices we tried, only the addition of $CH_3OH$ produces a sufficiently large improvement in $\chi^2$ to constitute a confident detection *(15)*. This does not preclude the presence of $H_2O$ or $NH_3$ ices, but the available data do not provide statistically significant evidence for their presence. Adding more than a trace of them to the models makes $\chi^2$ worse, but the increase can be minimized with large grain sizes that limit the projected area of $H_2O$ or $NH_3$ grains to a small fraction of the total. Small amounts of other ices, such as $H_2CO$, $CO_2$ or $C_2H_6$, could also be compatible with the data, as could the silicate and metallic phases seen in comets and interplanetary dust particles. The 1.8 µm feature in Arrokoth's spectrum is not matched by available ices or tholins and remains unidentified. It could be an artifact.

To search for spectral contrasts across the surface of Arrokoth, we selected several regions of interest (ROIs) as shown in Fig. 4B. These include LL, the brighter neck region (n), and a pair of ROIs on the left and right sides of SL (sl and sr), with sr incorporating the redder material on the rim of MD and sl representing portions of SL unrelated to MD. Spectra of these regions are shown in Fig. 4C. In averaging over fewer pixels, the signal/noise ratios in the ROI spectra are correspondingly poorer than the global average. However, the ROI spectra all look very similar to the average, and fitting Hapke models shows that tholin, carbon, and $CH_3OH$ are favored, without statistically significant evidence for $H_2O$ or $NH_3$ ices, just as with the average.

Our confidence in the detection of $CH_3OH$ is increased by the appearance of two distinct absorption bands of methanol ice. A band at 2.271 µm is attributed to ($v_1 + v_{11}$) or ($v_1 + v_7$) vibrational combination modes, and one at 2.338 µm is attributed to ($v_1 + v_8$) *(32)*. These spectral characteristics have been seen in Earth-based spectra of the Centaur 5145 Pholus *(33)* and the resonant KBO (55638) 2002 $VE_{95}$ *(34)*. Additional, weaker $CH_3OH$ absorption bands at 1.6 and 2.1 µm are not visible in Arrokoth's spectrum. In the case of Pholus, the spectrum from 0.45 to 2.45 µm was fitted with a radiative transfer model incorporating solid $CH_3OH$ and $H_2O$, in addition to an iron-bearing olivine (forsterite Fo 82) and tholin *(33)*, similar to our models for Arrokoth, except that $H_2O$ ice is not required in the Arrokoth models.

To assess spectral contrasts in a way that does not depend on multiple scattering models, a PCA was performed on the LEISA cube, with results shown in Fig. 5. Because of the low signal/noise ratio, we first binned the wavelengths down to 28 channels, producing an effective spectral resolving power of 39. We also discarded pixels along the edge where jitter during the scan is prone to producing artifacts. As with the MVIC colors, PC1 is sensitive to the overall light variation from shading and albedo, with the eigenvector flat across all wavelengths (Fig. 5A). PC1 accounts for 72% of the total variance in the LEISA data. PC2 captures only 2.3% of the variance, but the eigenvector shows pronounced dips around 1.5 and 2 µm, where $H_2O$ ice has its strongest absorption bands within LEISA's spectral range, suggesting that regional variations in $H_2O$ ice absorption could be the next most prominent source of spectral variance across the surface of Arrokoth, perhaps being most abundant around Maryland crater. However, the absence of strong $H_2O$ absorptions in any of our extracted spectra (including the sr ROI that covers this region) reduces confidence in that conclusion. Subsequent PCs account for even lower fractions of the total variance. The lack of spatial coherence in the images coupled with eigenvectors that are not suggestive of absorption by likely surface constituents, suggests that the higher PCs are responding mostly to instrumental noise rather than to signal in the LEISA data.

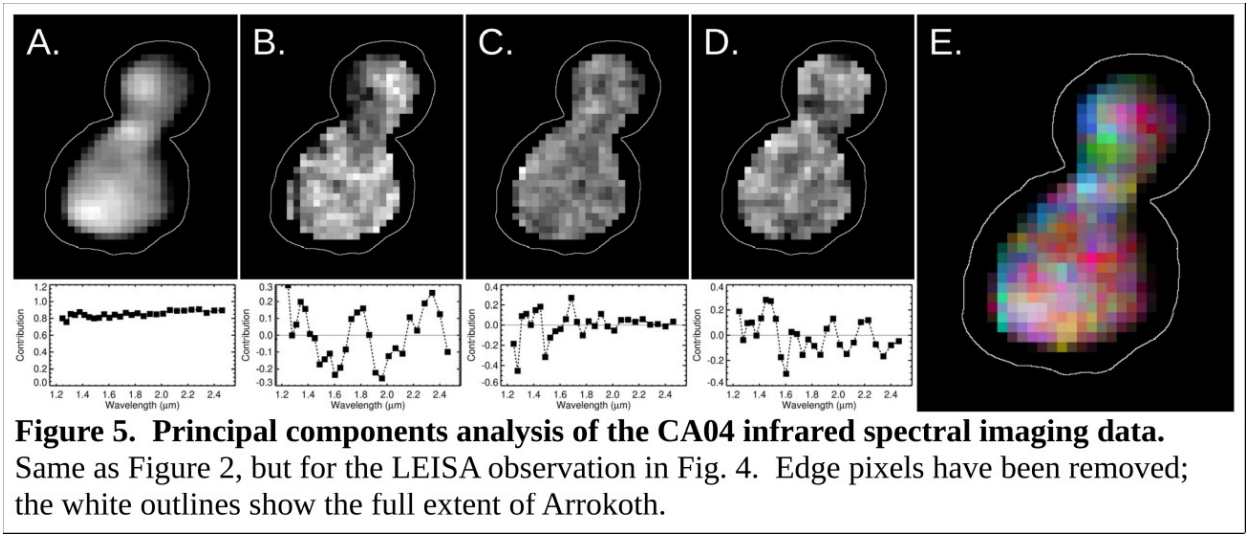

**Figure 5. Principal components analysis of the CA04 infrared spectral imaging data.** Same as Figure 2, but for the LEISA observation in Fig. 4. Edge pixels have been removed; the white outlines show the full extent of Arrokoth.

## Thermal Environment

The CA03 REX observation was performed on approach, observing Arrokoth's day side while the CA08 observation was done after closest approach, looking back at Arrokoth's night side. The microwave sky background is shown in Fig. 6A *(35, 36)*. The CA03 observation was performed with a fixed staring geometry. A later observation of the same field was obtained for background subtraction, but the system antenna temperature drifts over time, making it difficult to separate out the flux from Arrokoth. The CA08 night side observation was obtained under more favorable geometry, from a closer range, and with the antenna scanned along the uncertainty ellipse for Arrokoth's location. Scanning instead of staring facilitated calibration against the later background observation despite the drift in system response. The flux measurements are shown Fig. 6B. The radiometric signal was converted to radio brightness temperature using procedures developed earlier in the mission, accounting for the solid angle subtended by Arrokoth within the antenna beam *(15, 17, 36, 37)*. Accounting for the 414.4 km$^2$ cross section of Arrokoth and noise from instrument and background, we obtain a mean brightness temperature $T_B$, averaged across the night-side visible face of Arrokoth, of $T_B$ = 29 ± 5 K, which is within the range of brightness temperatures estimated from an earlier analysis *(1)*. To translate that brightness temperature to a kinetic temperature requires knowledge of the X-band emissivity of Arrokoth's surface, which is not known, but most likely lies in the range from 0.7 to 0.9 *(36)*.

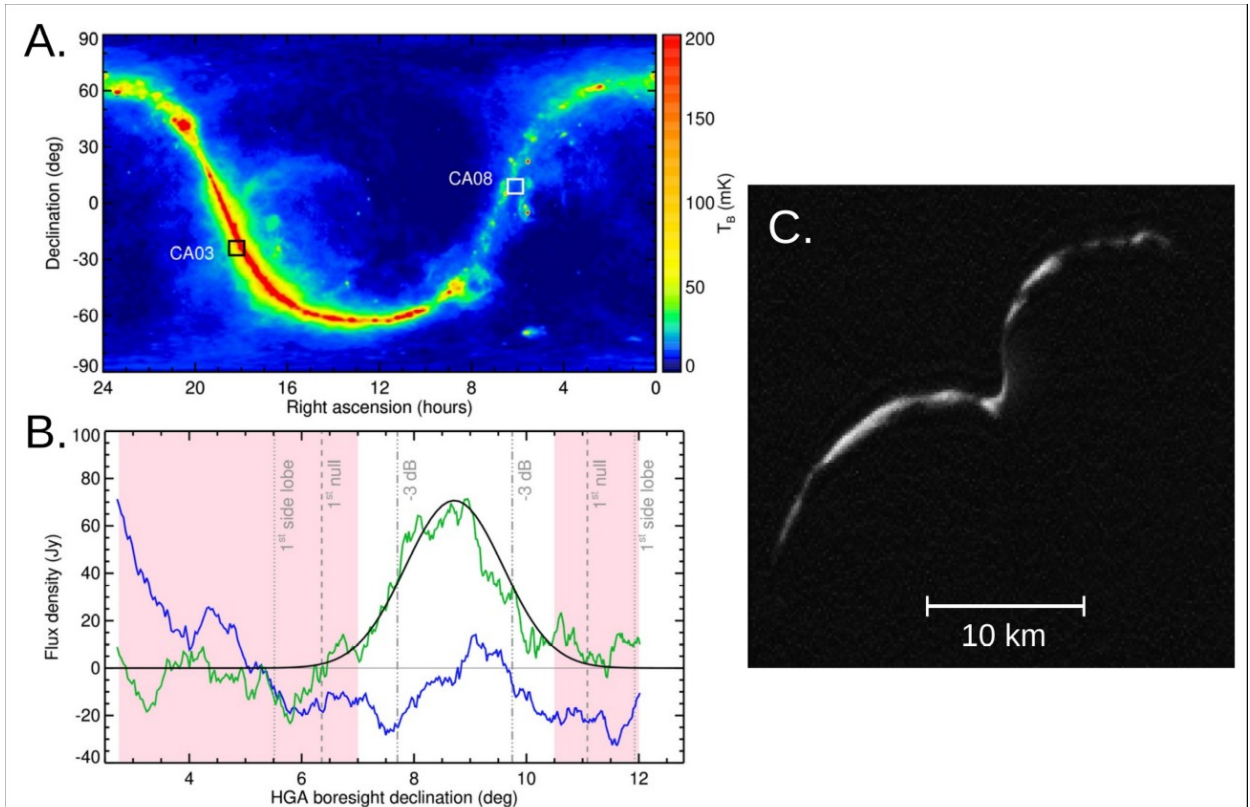

**Figure 6. Microwave radiometry of Arrokoth.** A. 7.1 GHz microwave sky background based on an all sky radio map *(35)* sampled at REX's 4.2 cm wavelength and smoothed to REX's 1.2° beam width *(36)*, indicating the locations of the CA03 and CA08 observations. B. Observed flux during the CA08 scan in green, with the later background scan in blue. Shaded areas were used to calibrate the two observations for background subtraction. The black curve is a model response for a $T_B$ = 30 K source with the 414 km$^2$ projected area of Arrokoth. C. CA07 LORRI image *(3)* obtained 10 minutes before the mid-time of the REX scan, at nearly identical lighting geometry but about a 10° shift in viewing geometry, showing more of the lit crescent than was visible at the time of the CA08 REX observation.

Thermophysical models were used to assess the implications of this $T_B$ measurement *(15)*. For each surface element of the three dimensional shape model *(3)*, we balanced radiative losses and thermal conduction against insolation (received sunlight) and also re-radiation from other parts of Arrokoth's surface visible from that location. Accounting for self-shadowing and surface re-radiation makes this modeling inherently global in scope *(38)*. Subsurface thermal evolution was simulated with a 1D thermal diffusion prescription *(15)*. For simplicity, we assume Arrokoth's obliquity does not precess and its orbit is circular, with a semimajor axis of 44.2 AU and period of 298 years. At this distance the incident solar radiation flux $F_\odot$ is 0.7 W m$^{-2}$. Given Arrokoth's 99.3° obliquity *(2)*, seasonal effects are strong. We determined the subsolar latitude along approximately 300 equally spaced temporal nodes over one orbital period *(15)*. During the New Horizons flyby the subsolar latitude was approximately −62°. At each orbital node, the daily averaged (15.9 hr period) solar insolation was calculated, accounting for self-shadowing. With these diurnally averaged insolation profiles, we determined the surface temperature on every element over the course of an orbit. We assumed that the subsurface thermal response is in the time-asymptotic limit, meaning there is no net gain or loss of thermal energy into or out of the interior over the course of one orbit *(39)*. This assumption requires heat from radioactive decay inside Arrokoth to be negligible and for the interior to have reached thermal equilibrium with the Sun over the course of the lifetime of the Solar System (see below).

We assume that the low bond albedo [$A_B$ = 0.06 *(1, 3)*] surface of Arrokoth is characterized by a very low thermal inertia ($\Gamma$ = 2.5 J m$^{-2}$ s$^{-1/2}$ K$^{-1}$) typical of loosely consolidated granular material, as inferred from infrared observations of KBOs *(40)*. The thermal inertia is given by $\Gamma \equiv \sqrt{k \rho C_p}$, where $k$ is the thermal conductivity, $\rho$ is the density, and $C_p$ is the specific heat at constant pressure. Arrokoth's bulk density must be at least 290 kg m$^{-3}$ *(3)* and densities of small KBOs and comet nuclei tend to be higher than that, but generally less than 1000 kg m$^{-3}$ *(41, 42)*. The density near the surface that matters for Arrokoth's thermal response is even more uncertain, and could differ substantially from the bulk density. Following *(3)*, we assume a generic density of $\rho$ = 500 kg m$^{-3}$ and that $C_p$ = 350 J kg$^{-1}$ K$^{-1}$. Under these assumptions, the corresponding thermal conductivity is 3.6×10$^{-5}$ W m$^{-1}$ K$^{-1}$, very low compared to values determined for surfaces in the inner Solar System. These values correspond to a seasonal thermal skin depth $\lambda$ = 0.55 m, where $\lambda \equiv \sqrt{k \tau_s (\rho C_p 2\pi)^{-1}}$, and $\tau_s$ is the 298 yr seasonal period *(43)*. This value is similar to the electrical skin depth *(36)*. The sub-solar equilibrium temperature $T_*$ = 58 K was obtained from $\epsilon \sigma T_*^4 = (1-A_B)F_\odot$, where $\epsilon$ is the emissivity, commonly assumed to be 0.9 [e.g., *(40, 43)*], $\sigma$ is the Stefan-Boltzmann constant (5.67×10$^{-8}$ W m$^{-2}$ K$^{-4}$), and $F_\odot$ is the solar flux at 44.2 AU. The thermal parameter $\Theta \equiv \Gamma \omega_0^{1/2} \sigma^{-1} T_*^{-3}$ characterizes the efficiency of the energy transport rate (per K) of thermal conduction across one seasonal skin depth relative to radiative losses *(43)*, which is about $\Theta \equiv 0.02$ for Arrokoth. Such a small value of $\Theta$ indicates that the surface layers are highly insulating, leading to extreme variations in surface temperature over the course of a year. Winter surface temperatures are much colder than peak summer temperatures. The low conductivity might only pertain to a surficial layer. If deeper below the surface the texture is more compacted with greater granular contact, the conductivity would likely be higher. We can estimate the body's thermalization timescale by calculating the thermal wave propagation time ($t_{\text{thermalization}} \equiv 2\pi \rho C_p R^2 k^{-1}$) across a length scale $R$ corresponding to the characteristic radius of the short axes of both lobes (~3.5 km). For values of $k$ greater than 10$^{-4}$ W m$^{-1}$ K$^{-1}$, this timescale is less than the age of the Solar System, supporting our assumption of a time-asymptotic state.

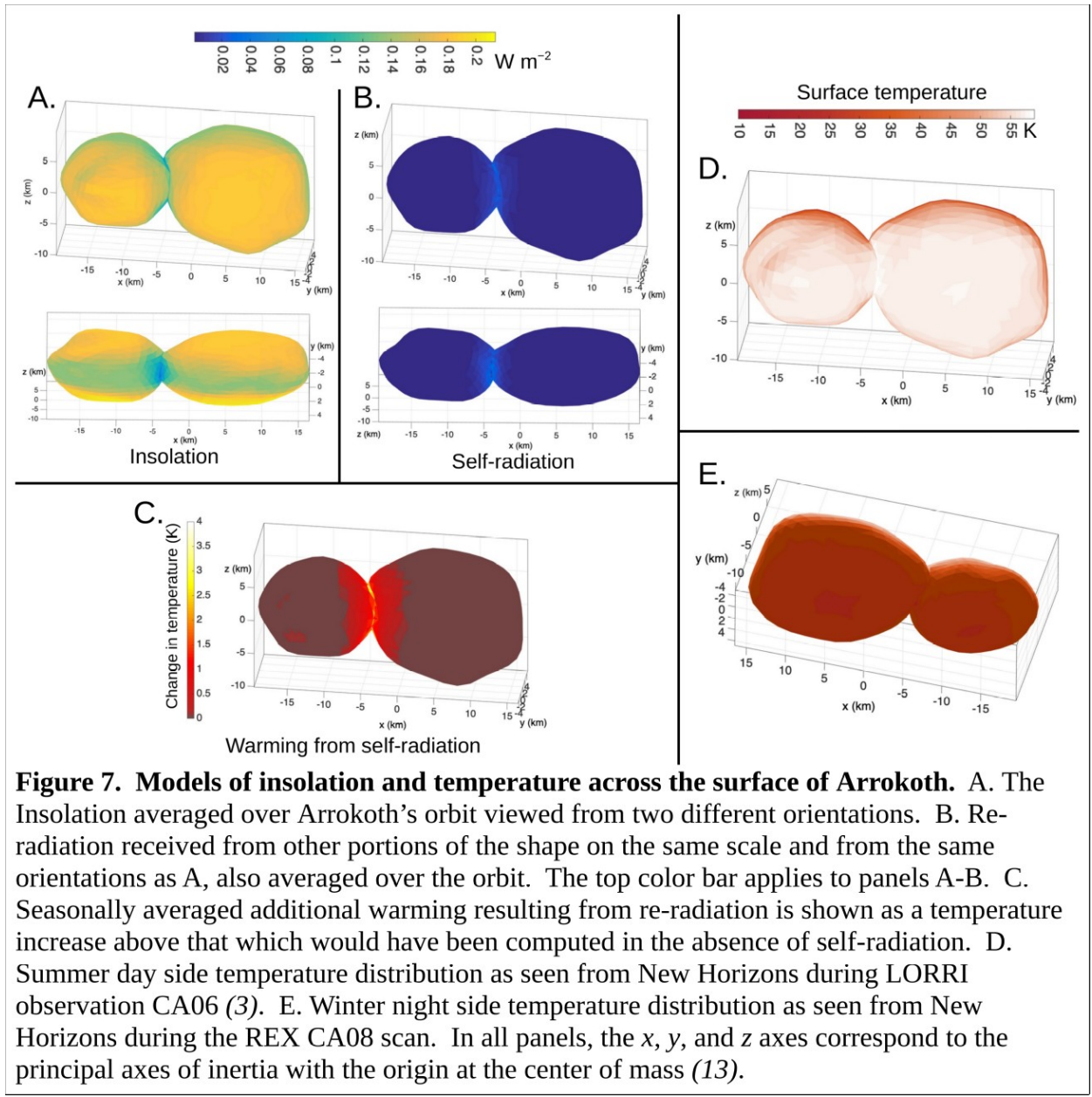

**Figure 7. Models of insolation and temperature across the surface of Arrokoth.** A. The Insolation averaged over Arrokoth's orbit viewed from two different orientations. B. Re-radiation received from other portions of the shape on the same scale and from the same orientations as A, also averaged over the orbit. The top color bar applies to panels A-B. C. Seasonally averaged additional warming resulting from re-radiation is shown as a temperature increase above that which would have been computed in the absence of self-radiation. D. Summer day side temperature distribution as seen from New Horizons during LORRI observation CA06 *(3)*. E. Winter night side temperature distribution as seen from New Horizons during the REX CA08 scan. In all panels, the *x*, *y*, and *z* axes correspond to the principal axes of inertia with the origin at the center of mass *(13)*.

Figure 7A shows the insolation averaged over an orbit from two viewing positions. The flattened shape and high obliquity lead to the equator receiving less energy on average (~0.1 W m$^{-2}$) than the poles (~0.2 W m$^{-2}$). Owing to self-shadowing, the neck region generally receives less energy than the equatorial zone (ranging from 0.06 to 0.08 W m$^{-2}$). Figure 7B shows the additional radiation received from thermal emission from other parts of Arrokoth itself, again averaged over an orbit. Our model indicates that the neck region is warmed by this trapping process, receiving about 0.025 to 0.04 W m$^{-2}$ from self-re-radiation, which partially offsets the effect of shadowing. Figure 7C shows the orbital average of the warming due to self-reradiation. The neck region experiences the greatest amount of relative warming, in the range of ~1-3 K. MD crater also receives enhanced thermal re-radiation, but the relative warming in that region is slight, about 0.5 K. Figure 7D shows the predicted observable surface temperature at the time of the New Horizons encounter. Typical temperatures are in the range of 50-57 K near the poles of each lobe, falling to ~40 K near the equator. Parts of the neck region that are not shaded at times of high subsolar latitude (−62° at the time of the encounter) stand out as having among the warmest surface temperatures during the encounter, as high as 60 K.

Figure 7E shows the predicted surface temperature at the time and viewing geometry of the CA08 REX observation, during which the sub spacecraft point was latitude +44°, longitude E 78°. From this orientation the model global averaged surface temperature is 16.1 K. While the surface temperature within the bulk of the body's winter night side is predicted to be in the range of 12 to 14 K, the contribution to the average from the viewable part of the lit equatorial region (top of Fig. 7E; $T$ ~ 40-55 K) raises the average temperature only slightly. The observed thermal emission seen by REX yields a much warmer mean brightness temperature of 29 K. If the model is correct, this discrepancy implies that the 4.2 cm radiation emerges primarily from the warmer subsurface, which is consistent with the expectation that the 4.2 cm thermal emission samples many wavelengths into the surface *(36)*. Higher thermal inertias than we have assumed would also permit warmer winter surface temperatures *(15)*.

## Implications for formation

The distribution of orbits in the present-day Kuiper belt was strongly influenced by an outward migration of Neptune early in Solar System history resulting from dynamical interaction between Neptune and the disk of planetesimals, e.g., *(44)*. Neptune's migration stopped at 30 AU from the Sun, indicating a break in the distribution of planetesimals in the disk beyond which there was insufficient mass to drive Neptune's migration further outward *(45)*. Most KBOs that have been studied spectroscopically [e.g., *(46, 47)*], are not CCKBOs; they originated in the denser planetesimal disk from inside 30 AU. Likewise, most comets that can be studied in detail due to their close approaches to the Sun and Earth, likely do not sample the outer planetesimal disk from beyond 30 AU. Arrokoth may contain a record of conditions in the outer part of the nebula where it formed. Constraints include the evidence for methanol ice and the lack of evidence for water ice, which is unlike the high abundance of $H_2O$ in many outer Solar System bodies and interstellar grains.

CCKBOs appear to have formed in the outer solar nebula via the gravitational collapse of pebble-size particles, concentrated aerodynamically *(13, 48)*. In this scenario, microscopic dust grains coagulate into larger particles *(49, 50)*. As particles approach pebble sizes they decouple from the gas, causing them to spend most of their time near the cold disk midplane and allowing them to become concentrated in dense clumps that can gravitationally collapse into planetesimals *(51, 52)*. The (original) bulk composition of CCKBOs should reflect the make-up of the solids present in the midplane of the solar nebula at the time and location of their formation. It remains unclear, however, how long the dust coagulation phase lasted and/or how far pebbles were able to move radially inward, e.g., *(53)*, before they formed planetesimals.

When the solar nebula formed, the chemical composition of the ices present in the outer regions was set by a combination of inheritance from the parent molecular cloud and chemistry taking place during formation of the disk. In the midplane of the outer disk, the resulting $CH_3OH/H_2O$ ratio on cold grain surfaces likely did not exceed a few percent *(54)*. During the subsequent disk evolution, spatial variations in physical conditions such as temperature, density, and radiation environment, coupled with ongoing chemistry *(55)* and transport/mixing processes *(56)*, result in gas-phase and ice compositions changing over time.

In regions where it was cold enough for highly-volatile CO to freeze as ice onto grains *(57)*, methanol could be formed through successive addition of hydrogen atoms to CO ice. Both interstellar and outer nebular environments are potential settings for this chemistry *(58, 59, 60)*. Before the loss of nebular gas and dust, the midplane of the disk was shaded from direct sunlight and extremely cold, favoring condensation of CO in its outermost portions. Simulations of protoplanetary disks indicate that methanol can be produced in this way (consuming CO) on ~Myr timescales at Arrokoth's distance from the Sun *(61)*. Intermediate steps include formation of formaldehyde ($H_2CO$) and radicals (e.g., $CH_3O$). Radiolytic destruction of $CH_3OH$ can produce $H_2CO$, but the band at 2.27 µm remains prominent even after irradiation *(62)*.

Another potential $CH_3OH$ formation mechanism involves radiolysis of mixed $H_2O$ and $CH_4$ ices *(63, 64, 65, 66)*. Again, low temperatures consistent with the shaded midplane are required for $CH_4$ to be frozen onto grains, although $CH_4$ is not quite as volatile as CO. If such radiolytic production occurs with an excess of $CH_4$, the $H_2O$ could be consumed, providing a possible explanation for the lack of evidence for $H_2O$ at Arrokoth. Such a radiolytic process would also efficiently form simple hydrocarbons such as $C_2H_4$ and $C_2H_6$, which are known to be precursors of complex organic tholins, e.g., *(20, 67)*.

Gas-phase methanol has been detected at low abundance in nearby protoplanetary disks *(68)*. This is consistent with methanol ice formation on grain surfaces, with a small fraction subsequently released to the gas through non-thermal desorption mechanisms, e.g., *(60, 69)*. This would also be consistent with the observation that many of these disks are also depleted in gaseous CO *(70, 71)*, requiring a combination of sequestration on pebbles and chemical processing *(61, 72, 73)*.

Although $H_2O$ was not detected on Arrokoth, it could be present, but somehow masked or hidden from view, such as by materials produced through radiolysis or photolysis of $CH_3OH$ ice and perhaps other, undetected precursor materials. Preferential removal of $H_2O$ ice from the uppermost surface by a process such as sputtering is another possibility, though it is unclear that $H_2O$ should be more susceptible to such removal than $CH_3OH$ is. Spectra of some larger KBOs also lack $H_2O$ absorption features *(46, 47)*. $H_2O$ ice absorption is considerably weaker in the spectrum of Arrokoth than seen on Pholus and (55638) 2002 $VE_{95}$, the two other objects with strong $CH_3OH$ signatures, but those objects are much larger than Arrokoth *(74, 75)*. They likely formed in the closer, more densely populated planetesimal disk originally inside 30 AU, as did other, large KBOs where strong $H_2O$ ice signatures have been detected spectroscopically. It is hard to envision a mechanism that preferentially masks or removes $H_2O$ from the surface of Arrokoth but not the $H_2O$ on these other objects. Their contrasting compositions suggests that the observed surface composition of these bodies is reflective of their bulk compositions, and Arrokoth's composition is distinct from those of planetesimals that formed closer to the Sun. A contrast in planetesimal composition driven by nebular chemistry enabled by CO and/or $CH_4$ frozen on grains may be connected to the transition at 30 AU that halted Neptune's outward migration at that distance.

Although regional variations in tholin and ice abundance could cause albedo, color, and spectral variations, the subtle variations that are seen at Arrokoth do not require compositional differences. Reflectance also depends on mechanical properties such as the particle size distribution and degree of compaction *(31)*. The merger of the two lobes *(13)* could have mechanically modified the material in the neck region. After the formation of Arrokoth from the nebula, low speed impacts of residual debris could locally modify surface textures, which might account for some of the spots with slightly contrasting colors and albedos.

## Surface and interior evolution

The surface features of comets are dominated by geologically rapid volatile loss and sublimation erosion, whilst the surfaces of larger asteroids are dominated by high-energy impacts. In contrast, Arrokoth and the CCKBOs are distinct in inhabiting an environment with very little energy input from interstellar, solar, and micrometeorite sources that require long timescales to modify the surface. Depending on the thermal parameters, surface temperatures range from as low as 10 to 20 K in winter to 50 to 60 K in summer, with the neck region getting at most a few degrees warmer due to self-radiation. Summer surface temperatures are warm enough to drive off volatiles such as CO, $CH_4$, and $N_2$, but are not warm enough to crystallize amorphous $H_2O$ ice, or to sublimate it. We therefore expect little thermally-driven evolution of the surface, except early in Arrokoth's history when the volatile ices would have been lost soon after nebular dust cleared, allowing sunlight to illuminate the surface. Galactic and solar energetic photons and charged particles can break bonds and drive chemical reactions that produce refractory macromolecular tholins *(19, 24, 25)*. Photolysis and radiolysis of solid methanol also produces formaldehyde, which may subsequently polymerize *(76, 77)*. While formaldehyde polymer (paraformaldehyde) shows some spectral structure in the 2-2.5 μm region, it does not match the bands seen in Pholus *(33)* or Arrokoth. Macromolecular tholins seen on Arrokoth's surface could derive from three potential sources. (*i*) They could have originated in the pre-solar gas and dust cloud. (*ii*) They could arise from photolysis and radiolysis of hydrocarbons and associated nitrogen- and oxygen-bearing components in the nebula, especially where material is transported to regions near the surface of the nebula exposed to radiation from the forming Sun or other astrophysical sources *(56)*. (*iii*) Radiolysis and photolysis of Arrokoth's surface components could produce such tholins, as discussed above. All of these sources likely contributed to Arrokoth's inventory of complex organics, with the products of the first two mechanisms being distributed throughout the body, while the products of the third only occurring at the surface. The flux of Solar System and interstellar micrometeorites at Arrokoth's location is highly uncertain *(78, 79)*, but such bombardment could produce up to several meters of mechanical erosion over the age of the Solar System *(80, 81)*. If this erosion operates faster than space weathering by energetic radiation, the visible surface could be representative of the deep interior. If it is slower, the surface should accumulate a lag deposit enriched in more refractory materials through loss or destruction of more volatile and fragile molecules.

It is not obvious from the encounter data that a distinct surface veneer exists on Arrokoth. Albedo and color contrasts corresponding to ancient features such as the neck suggest that such contrasts are not quickly masked by a space weathering processes. If compositionally distinct interior material was exposed at the neck, it might weather differently and thus maintain a contrasting appearance, but the LEISA data show no evidence for a distinct composition in the neck region. The warmer thermal environment of the neck would be another potential reason for distinct evolution there, but the temperature difference is too small to produce outcomes that differ substantially from the rest of the surface. No obviously fresh craters expose distinct-looking interior material (with the possible exception of Maryland), and color trends do not appear to correspond to down-slope transport, which is generally from equators to the poles of the lobes, and ultimately to the neck *(3)*. Less-altered interior material should be preferentially exposed at high elevations, but we do not see obvious color differences in high-standing regions along the equators. Brighter material in the neck and in Maryland may accumulate in topographic lows, suggestive of textural rather than compositional contrasts. Among the CCKBO population, the diversity of colors coupled with similar colors of different sized components of binaries have been used to argue against the importance of size-dependent factors such as the balance between erosion and space weathering in altering their surface colors *(82)*.

The evolution of Arrokoth's interior is shaped by even lower energy inputs than its surface. Subsequent to the loss of shading from nebular dust, insolation would have raised Arrokoth's equilibrium temperature. As that thermal wave slowly propagated inward, highly volatile species such as $N_2$, CO, and $CH_4$ would have become unstable, at least as condensed ices. Early outgassing of these species may have produced what appear to be collapse or outgassing pits at the boundaries of terrain units *(1, 3)*. Such features may be analogous to pits or sinkholes on comet 67P/Churyumov-Gerasimenko, e.g., *(83, 84)*. Localization of the pits to certain regions may arise from variable permeability of surface deposits that would favor volatiles escaping through weaker zones at unit boundaries. However the equilibrium temperature in Arrokoth's interior would never have been high enough for amorphous ice to crystallize and expel its payload of trapped volatiles, e.g., *(85, 86)*. Apart from loss of these volatile species, Arrokoth's interior may have undergone little alteration or processing since accretion, and could thus preserve many characteristics of the original accretion such as layering [as observed on comets *(87, 88)*], very high porosity, and an intimate mixture of nebular ices, organics, and silicate dust grains.

# References and Notes

1. S. A. Stern et al., Initial results from the New Horizons exploration of 2014 MU$_{69}$, a small Kuiper Belt object. *Science* **364,** eaaw9771 (2019).

2. S. B. Porter et al., High-precision orbit fitting and uncertainty analysis of (486958) 2014 MU$_{69}$. *Astron. J.* **156,** 20 (2018).

3. J. R. Spencer et al., The geology and geophysics of Kuiper belt object (486958) Arrokoth. *Science* (this issue; 2020) DOI:10.1126/science.aay3999.

4. J.-M. Petit et al., The Canada-France Ecliptic Plane Survey - full data release: The orbital structure of the Kuiper belt. *Astron. J.* **142,** 131 (2011).

5. H.F. Levison, A. Morbidelli, C. Van Laerhoven, R. Gomes, K. Tsiganis, Origin of the structure of the Kuiper belt during a dynamical instability in the orbits of Uranus and Neptune. *Icarus* **196,** 258-273 (2008).

6. K. S. Noll, W. M. Grundy, D. C. Stephens, H. F. Levison, S. D. Kern, Evidence for two populations of classical transneptunian objects: The strong inclination dependence of classical binaries. *Icarus* **194,** 758-768 (2008).

7. S. C. Tegler, W. Romanishin, Extremely red Kuiper-belt objects in near-circular orbits beyond 40 AU. *Nature* **407,** 979-981 (2000).

8. R. E. Pike et al., Col-OSSOS: z-band photometry reveals three distinct TNO surface types. *Astron. J.* **154,** 101 (2017).

9. H. F. Levison, S. A. Stern, On the size dependence of the inclination distribution of the main Kuiper Belt. *Astron. J.* **121,** 1730-1735 (2001).

10. M. E. Schwamb, M. E., Brown, W. C. Fraser, The small numbers of large Kuiper belt objects. *Astron. J.* **147,** 2 (2014).

11. M. J. Brucker et al., High albedos of low inclination Classical Kuiper belt objects. *Icarus* **201,** 284-294 (2009).

12. E. Vilenius et al., "TNOs are Cool": A survey of the trans-neptunian region X. Analysis of classical Kuiper belt objects from Herschel and Spitzer observations. *Astron. & Astrophys.* **564,** A35 (2014).

13. W. B. McKinnon et al., The solar nebula origin of (486958) Arrokoth, a primordial contact binary in the Kuiper Belt. *Science* (this issue; 2020) DOI:10.1126/science.aay6620.

14. D. C. Reuter et al., Ralph: A visible/infrared imager for the New Horizons Pluto/Kuiper belt mission. *Space Sci. Rev.* **140,** 129-154 (2008).

15. Materials and methods are available as supplementary materials.

16. A. F. Cheng et al., LOng-Range Reconnaissance Imager on New Horizons. *Space Sci. Rev.* **140,** 189-215 (2008).

17. G. L. Tyler et al., The New Horizons radio science experiment (REX). *Space Sci. Rev.* **140,** 217-259 (2008).

18. O. R. Hainaut, A. C. Delsanti, Colors of minor bodies in the outer Solar System: A statistical analysis. *Astron. & Astrophys.* **389,** 641-664 (2002).

19. D. P. Cruikshank, H. Imanaka, C. M. Dalle Ore, Tholins as coloring agents on outer Solar System bodies. *Adv. Space Res.* **36,** 178-183 (2005).

20. B. N. Khare, C. Sagan, Optical constants of organic tholins produced in a simulated Titanian atmosphere: From soft X-ray to microwave frequencies. *Icarus* **60,** 127-134 (1984).

21. B. N. Khare et al., Production and optical constants of ice tholin from charged particle irradiation of (1:6) $C_2H_6/H_2O$ at 77 K. *Icarus* **103,** 290-300 (1993).

22. B. N. Khare et al., Optical constants of Triton tholin: Prelimary results. *Bull. Am. Astron. Soc.* **26,** 1176-1177 (1994).

23. H. Imanaka et al., Laboratory experiments of Titan tholin formed in cold plasma at various pressures: Implications for nitrogen-containing polycyclic aromatic compounds in Titan haze. *Icarus* **168,** 344-366 (2004).

24. C. K. Materese et al., Ice chemistry on outer Solar System bodies: Carboxylic acids, nitriles, and urea detected in refractory residues produced from the UV photolysis of $N_2$:$CH_4$:CO-containing ices. *Astrophys. J.* **788,** 111 (2014).

25. C. K. Materese, D. P. Cruikshank, S. A. Sandford, H,. Imanaka, M. Nuevo, Ice chemistry on outer Solar System bodies: Electron radiolysis of $N_2$-, $CH_4$-, and CO-containing ices. *Astrophys. J.* **812,** 150 (2015).

26. N. Peixinho, A. Delsanti, A. Guilbert-Lepoutre, R. Gafeira, P. Lacerda, The bimodal colors of Centaurs and small Kuiper belt objects. *Astron. & Astrophys.* **546,** A86 (2012).

27. F. E. DeMeo, B. Carry, The taxonomic distribution of asteroids from multi-filter all-sky photometric surveys. *Icarus* **226,** 723-741 (2013).

28. G. M. Szabó, Z. Ivezić, M. Jurić, R. Lupton, The properties of jovian Trojan asteroids listed in SDSS Moving Object Catalogue 3. *Mon. Not. R. Astron. Soc.* **377,** 1393-1406 (2007).

29. W. C. Fraser, M. E. Brown, The Hubble Wide Field Camera 3 test of surfaces in the outer Solar System: The compositional classes of the Kuiper belt. *Astrophys. J.* **749,** 33 (2012).

30. B. Schmitt et al., Physical state and distribution of materials at the surface of Pluto from New Horizons LEISA imaging spectrometer. *Icarus* **287,** 229-260 (2017).

31. B. Hapke, *Theory of reflectance and emittance spectroscopy, 2nd edition*. Cambridge University Press, New York (2012).

32. S. A. Sandford, L. J. Allamandola, Condensation and vaporization studies of $CH_3OH$ and $NH_3$ ices: Major implications for astrochemistry. *Astrophys. J.* **417,** 815-825 (1993).

33. D. P. Cruikshank et al., The composition of Centaur 5145 Pholus. *Icarus* **135,** 389-407 (1998).

34. M. A. Barucci, F. Merlin, E. Dotto, A. Doressoundiram, C. de Bergh, TNO surface ices: Observations of the TNO 55638 (2002 $VE_{95}$) and analysis of the population's spectral properties. *Astron. & Astrophys.* **455,** 725-730 (2006).

35. H. Zheng et al., An improved model of diffuse galactic radio emission from 10 MHz to 5 THz. *Mon. Not. R. Astron. Soc.* **464,** 3486-3497 (2017).

36. M. K. Bird et al., Radio thermal emission from Pluto and Charon during the New Horizons encounter. *Icarus* **322,** 192-209 (2019).

37. I. R. Linscott et al., Radiometer Calibration at 4.2 cm on New Horizons. Stanford Radioscience Report No. 17-06-0, available at https://sbn.astro.umd.edu/holdings/nh-j-rex-2-jupiter-v2.0/document/nh_rex_radiometer_calib_v4p7.pdf (2017).

38. D. Vokrouhlický, W. F. Bottke, S. R. Chesley, D. J. Scheeres, T. S. Statler, The Yarkovsky and YORP effects. In *Asteroids IV*, P. Michel, F. E. DeMeo, W. F. Bottke (Eds.), University of Arizona Press, Tucson, pp. 509-531 (2015).

39. O. L. White, O. M. Umurhan, J. W. Moore, A. D. Howard, Modeling of ice pinnacle formation on Callisto. *J. Geophys. Res.* **121,** 21-45 (2016).

40. E. Lellouch et al., "TNOs are Cool": A survey of the trans-neptunian region IX. Thermal properties of Kuiper belt objects and Centaurs from combined Herschel and Spitzer observations. *Astron. & Astrophys.* **557,** A60 (2013).

41. M. Pätzold et al., A homogeneous nucleus for comet 67P/Churyumov-Gerasimenko from its gravity field. *Nature* **530,** 63-65 (2016).

42. C. J. Bierson, F. Nimmo, Using the density of Kuiper belt objects to constrain their composition and formation history. Icarus 326, 10-17 (2019).

43. J. R. Spencer, A rough-surface thermophysical model for airless planets. *Icarus* **83,** 27-38 (1990).

44. R. S. Gomes, A. Morbidelli, H. F. Levison, Planetary migration in a planetesimal disk: Why did Neptune stop at 30 AU? *Icarus* **170,** 492-507 (2004).

45. A. Shannon, Y. Wu, and Y. Lithwick, Forming the cold classical Kuiper belt in a light disk. *Astrophys. J.* **818,** 175 (2016).

46. K. M. Barkume, M. E. Brown, and E. L. Schaller, Near-infrared spectra of Centaurs and Kuiper belt objects. *Astron. J.* **135,** 55-67 (2008).

47. M. A. Barucci et al., New insights on ices in Centaur and transneptunian populations. *Icarus* **214,** 297-307 (2011).

48. D. Nesvorný, R. Li, A. Youdin, J. B. Simon, W. M. Grundy, Trans-neptunian binaries as evidence for planetesimal formation by the streaming instability. *Nature Astron.,* 349 (2019).

49. C. Dominik, A. G. G. M. Tielens, The physics of dust coagulation and the structure of dust aggregates in space. *Astrophys. J.* **480,** 647-673 (1997).

50. J. Blum, G. Wurm, The growth mechanisms of macroscopic bodies in protoplanetary disks. *Annu. Rev. Astron. Astrophys.* **46,** 21-56 (2008).

51. A. N. Youdin, J. Goodman, Streaming instabilities in protoplanetary disks. *Astrophys. J.* **620,** 459-469 (2005).

52. A. Johansen et al., Rapid planetesimal formation in turbulent circumstellar disks. *Nature* **448,** 1022-1025 (2007).

53. S. J. Weidenschilling, Aerodynamics of solid bodies in the solar nebula. *Mon. Not. R. Astron. Soc.* **180,** 57-70 (1977).

54. M. N. Drozdovskaya et al., Cometary ices in forming protoplanetary disc midplanes. *Mon. Not. R. Astron. Soc.* **462,** 977-993 (2016).

55. T. Henning, D. Semenov, Chemistry in protoplanetary disks. *Chem. Rev.* **113,** 9016-9042 (2013).

56. F. J. Ciesla, S. A. Sandford, Organic synthesis via irradiation and warming of ice grains in the solar nebula. *Science* **336,** 452 (2012).

57. K. I. Öberg, R. Murray-Clay, E. A. Bergin, The effect of snowlines on C/O in planetary atmospheres. *Astrophys. J.* **743,** L16 (2011).

58. A. G. G. M. Tielens, W. Hagen, Model calculations of the molecular composition of interstellar grain mantles. *Astron. & Astrophys.* **114,** 245-260 (1982).

59. H. M. Cuppen, E. F. van Dishoeck, E. Herbst, A. G. G. M. Tielens, Microscopic simulation of methanol and formaldehyde ice formation in cold dense cores. *Astron. & Astrophys.* **508,** 275-287 (2009).

60. M. Ruaud, U. Gorti, A three-phase approach to grain surface chemistry in protoplanetary disks: Gas, ice surfaces, and ice mantles of dust grains. *Astrophys. J.* **885,** 146 (2019).

61. A. D. Bosman, C. Walsh, E. F. van Dishoeck, CO destruction in protoplanetary disk midplanes: Inside versus outside the CO snow surface. *Astron. & Astrophys.* **618,** A182 (2018).

62. R. Brunetto, G. A. Baratta, M. Domingo, G. Strazzulla, Reflectance and transmittance spectra (2.2-2.4 μm) of ion irradiated methanol. *Icarus* **175,** 226-232 (2005).

63. M. H. Moore, R. L. Hudson, Infrared study of ion-irradiated water-ice mixtures with hydrocarbons relevant to comets. *Icarus* **135,** 518-527 (1998).

64. A. Wada, N. Mochizuki, K. Hiraoka, Methanol formation from electron-irradiated mixed $H_2O/CH_4$ ice at 10 K. *Astrophys. J.* **644,** 300-306 (2006).

65. R. Hodyss, P. V. Johnson, J. V. Stern, J. D. Goguen, I. Kanik, Photochemistry of methane–water ices. *Icarus* **200,** 338-342 (2009).

66. M. P. Pearce et al., Formation of methanol from methane and water in an electrical discharge. *Phys. Chem. Chem. Phys.* **14,** 3444-3449 (2012).

67. G. D. McDonald et al., Production and chemical analysis of cometary ice tholins. *Icarus* **122,** 107-117 (1996).

68. C. Walsh et al., First detection of gas-phase methanol in a protoplanetary disk. *Astrophys. J.* **823,** L10 (2016).

69. C. Walsh, S. Vissapragada, H. McGee, Methanol formation in TW Hya and future prospects for detecting larger complex molecules in disks with ALMA. In *Astrochemistry VII - Through the Cosmos from Galaxies to Planets Proceedings IAU Symposium No. 332*, M. Cunningham, T. Millar, Y. Aikawa (Eds.), Proc. International Astronomical Union, pp. 395-402 (2018).

70. E. A. Bergin et al., An old disk still capable of forming a planetary system. *Nature* **493,** 644-646 (2013).

71. M. K. McClure et al., Mass measurements in protoplanetary disks from hydrogen deuteride. *Astrophys. J.* **831,** 167 (2016).

72. S. Krijt, K. R. Schwarz, E. A. Bergin, F. J. Ciesla, Transport of CO in protoplanetary disks: Consequences of pebble formation, settling, and radial drift. *Astrophys. J.* **864,** 78 (2018).

73. K. Zhang, E. A. Bergin, K. Schwarz, S. Krijt, F. Ciesla, Systematic variations of CO gas abundance with radius in gas-rich protoplanetary disks. *Astrophys. J.* **883,** 98 (2019).

74. S. C. Tegler et al., The period of rotation, shape, density, and homogeneous surface color of the Centaur 5145 Pholus. *Icarus* **175,** 390-396 (2005).

75. M. A. Barucci et al., The extra red plutino (55638) 2002 $VE_{95}$. *Astron. & Astrophys.* **539,** A152 (2012).

76. M. P. Bernstein, S. A. Sandford, L. J. Allamandola, S. Chang, M. A. Scharberg, Organic compounds produced by photolysis of realistic interstellar and cometary ice analogs containing methanol. *Astrophys. J.* **454,** 327-344 (1995).

77. T. Butscher et al., Radical-assisted polymerization in interstellar ice analogues: Formyl radical and polyoxymethylene. *Mon. Not. R. Astron. Soc.* **486,** 1953-1963 (2019).

78. M. Piquette et al., Student Dust Counter: Status report at 38 AU. *Icarus* **321,** 116-125 (2019).

79. A. R. Poppe et al., Constraining the Solar System's debris disk with in situ New Horizons measurements from the Edgeworth-Kuiper belt. *Astrophys. J. Lett.* **881,** L12.

80. S. A. Stern, ISM-induced erosion and gas-dynamical drag in the Oort cloud. *Icarus* **84,** 447-466 (1990).

81. S. A. Stern, The evolution of comets in the Oort cloud and Kuiper belt. *Nature* **424,** 639-642 (2003).

82. S. D. Benecchi et al., The correlated colors of transneptunian binaries. *Icarus* **200,** 292-303 (2009).

83. O. Mousis et al., Pits formation from volatile outgassing on 67P/Churyumov-Gerasimenko. *Astrophys. J. Lett.* **814,** L5 (2015).

84. J. B. Vincent et al., Large heterogeneities in comet 67P as revealed by active pits from sinkhole collapse. *Nature* **523,** 63-68 (2015).

85. P. Ayotte et al., Effect of porosity on the adsorption, desorption, trapping, and release of volatile gases by amorphous solid water. *J. Geophys. Res.* **106,** 33387-33392 (2001).

86. A. Bar-Nun, G. Notesco, T. Owen, Trapping of $N_2$, CO and Ar in amorphous ice: Application to comets. *Icarus* **190,** 655-659 (2007).

87. M. Massironi et al., Two independent and primitive envelopes of the bilobate nucleus of comet 67P. *Nature* **526,** 402-405 (2015).

88. L. Panasa et al., A three-dimensional modelling of the layered structure of comet 67P/Churyumov-Gerasimenko. *Mon. Not. R. Astron. Soc.* **469,** S741-S754 (2017).

89. C. J. A. Howett et al., Inflight radiometric calibration of New Horizons' Multispectral Visible Imaging Camera (MVIC). *Icarus* **287,** 140-151 (2017).

90. C. M. Dalle Ore et al., Ices on Charon: Distribution of $H_2O$ and $NH_3$ from New Horizons LEISA observations. *Icarus* **300,** 21-32 (2018).

91. C. H. Acton, Ancillary data services of NASA's Navigation and Ancillary Information Facility. *Planet. & Space Sci.* **44,** 65-70 (1996).

92. J. C. Cook et al., The distribution of $H_2O$, $CH_3OH$, and hydrocarbon-ices on Pluto: Analysis of New Horizons spectral images. *Icarus* **331,** 148-169 (2019).

93. A. F. McGuire, B. W. Hapke, An experimental study of light scattering by large, irregular particles. *Icarus* **113,** 134-155 (1995).

94. F. Rouleau, P. G. Martin, Shape and clustering effects on the optical properties of amorphous carbon. *Astrophys. J.* **377,** 526-540 (1991).

95. W. M. Grundy, B. Schmitt, The temperature-dependent near-infrared absorption spectrum of hexagonal $H_2O$ ice. *J. Geophys. Res.* **103,** 25809-25822 (1998).

96. A. Zanchet et al., Optical constants of $NH_3$ and $NH_3$:$N_2$ amorphous ices in the near-infrared and mid-infrared regions. *Astrophys. J.* **777,** 26 (2013).

97. T. N. Titus, G. E. Cushing, Thermal diffusivity experiment at the Grand Falls dune field. Third International Planetary Dunes Workshop, June 12-15 in Flagstaff, Arizona. LPI Contribution No. 7044 (2012).

98. A. P. Ingersoll, T. Svitek, B. C. Murray, Stability of polar frosts in spherical bowl-shaped craters on the moon, Mercury, and Mars. *Icarus* **100,** 40-47 (1992).

# Acknowledgments:

We gratefully thank the many hundreds of people who worked together on the New Horizons team. Their hard work enabled the encounter. **Funding:** Supported by NASA's New Horizons project via contracts NASW-02008 and NAS5-97271/TaskOrder30. JJK was supported by the National Research Council of Canada. SK acknowledges the support of a Hubble Fellowship, program number HST-HF2-51394.002 provided by NASA through a grant from the Space Telescope Science Institute, which is operated by the Association of Universities for Research in Astronomy, Incorporated, under NASA contract NAS5-26555. MR acknowledges support by the NASA Astrobiology Institute under Cooperative Agreement Notice NNH13ZDA017C issued through the Science Mission Directorate. BS, LG, and EQ acknowledge support by the French Centre National d'Etudes Spatiales (CNES). OMU acknowledges NASA Astrophysics Theory Program Grant Number NNX17AK59G for partial support of this work. **Author contributions:** WMG led authorship of the manuscript. WMG, CJAH, CBO, AHP, and SP analyzed the MVIC color data and wrote that section. WMG, AHP, SP, JCC, and DPC analyzed the LEISA spectral data and wrote that section. MKB and IRL analyzed the REX data and wrote that section. OMU did the thermal modeling and wrote that section, with input from WMG, CJAH, and LAY. WMG, DPC, DTB, SK, and MR wrote the implications section. CMDO, JJK, JTK, YJP, SBP, FS, JRS, SAS, AJV, and HAW contributed to data analysis and development of scientific ideas. All authors played roles in experimental design and/or provided feedback and insights on drafts of the manuscript. **Competing interests:** We declare no competing interests. **Data and materials availability:** All images, spacecraft data, and the shape model used in this paper are available at https://doi.org/10.6084/m9.figshare.c.4819110 and at https://doi.org/10.6084/m9.figshare.11485443. The three Charon observations used in flat-fielding the LEISA data are available at https://pds-smallbodies.astro.umd.edu/holdings/nh-p-leisa-3-pluto-v3.0/data/20150714_029914/lsb_0299146219_0x54b_sci.fit, https://pds-smallbodies.astro.umd.edu/holdings/nh-p-leisa-3-pluto-v3.0/data/20150714_029917/lsb_0299171308_0x53c_sci.fit, and https://pds-smallbodies.astro.umd.edu/holdings/nh-p-leisa-3-pluto-v3.0/data/20150714_029917/lsb_0299175509_0x53c_sci.fit. Additional fully calibrated New Horizons data and higher-order data products will be released by the NASA Planetary Data System at https://pds-smallbodies.astro.umd.edu/data_sb/missions/newhorizons/index.shtml in a series of stages in 2020 and 2021, due to the time required to fully downlink and calibrate the data.

## Supplementary Material

Materials and Methods

Ralph Instrument Details and Data Processing

The Ralph instrument is a three-mirror anastigmat telescope with a dichroic beamsplitter that feeds two focal planes, making it effectively two instruments in one *(14)*. The MVIC focal plane consists of 6 time delay integration (TDI) charge-coupled device (CCD) arrays and one frame transfer CCD. The TDI detectors have 32 rows to build up the integrated signal as the instrument scans the field of view across the scene. In the cross-track direction, orthogonal to the 32 TDI lines, there are 5000 active pixels. Each pixel subtends 20 µrad for a cross-track field of view of 5.7°. Ralph/MVIC is distinguished from other New Horizons imaging instruments for its wide field of view and color imaging capabilities. Four of the 6 TDI detectors have broad-band filters fixed to the detector assembly. The color filters are identified as "BLUE" (400-550 nm), "RED" (540-700 nm), "NIR" (780-975 nm) and "CH4" (860-910 nm). The two Pan-chromatic TDI detectors cover the wavelength range from 400-975 nm; see *(89)* for details of system response and calibration. Comparing the observed signal for a well lit portion of Arrokoth to the standard deviation in a sample of blank sky, we estimate the single pixel signal-to-noise ratios to be 60, 175, 148, and 57 in BLUE, RED, NIR, and CH4 filters, respectively. The corresponding 1-σ uncertainty in a percent color slope measurement is about ±0.2 around the mean slope of 27%. The uncertainty in determination of a mean color slope drops rapidly as multiple pixels are averaged together.

Ralph's LEISA focal plane consists of a Rockwell Scientific Corporation PICNIC array (a 256×256 pixel HgCdTe device) with two linear variable filters to disperse the light. The larger filter covers wavelengths from 1.25 to 2.5 µm with a resolving power ($\lambda/\Delta\lambda$, where $\lambda$ is the wavelength) of ~240. The smaller filter covers 2.1 to 2.25 µm with a higher resolving power (~560).

The LEISA detector has a field of view of 0.9° with both axes providing spatial data coupled with spectral dispersion along one axis.

For all the data collecting modes of Ralph (except observing with the panchromatic frame transfer array), the spacecraft scans the Ralph boresight across the target to build up an image or image cube in the case of LEISA. At Arrokoth, our scan rates were slower than previously used at Pluto because of the low light levels at Arrokoth's extreme distance from the Sun (43 au at the time of the encounter) and its low albedo. The color scans used a commanded rate of approximately 800 µrad s$^{-1}$ which corresponds to an integration time of 0.8 seconds. The instrument reads the actual scan rate of the spacecraft to set the TDI rate or frame rate for data collection: the actual scan rate is within ±32 µrad s$^{-1}$ of the commanded scan rate. For the LEISA observations, we used a very slow scan rate (60 µrad s$^{-1}$) to maximize the observed signal.

We used three LEISA scans of Charon taken during the Pluto encounter to construct a flat field for LEISA. Charon was used for this purpose because of the lack of spectral diversity across its surface, unlike Pluto. The three highest spatial resolution scans of Charon *(90)* were used in order to check for consistency. These are identified as C_LEISA_HIRES, C_LEISA_LORRI and C_LEISA. In each of these scans, Charon passed through the center of the field-of-view with the C_LEISA_HIRES scan subtending the greatest area of the detector. Charon did not completely fill the detector in any of these scans, so there are regions around the frame edges which have not been corrected, but Arrokoth was not imaged in these uncorrected regions.

Using the frame data (spatial vs. spatial-spectral format), a mask was constructed to isolate Charon from the background. The average signal was then calculated for each pixel when it was illuminated by Charon. This produced a 256×256 pixel array (with pixels set to zero if never illuminated by Charon) with the mean spectrum of Charon still present. That was removed by robust fitting (i.e., with rejection of outlier points) of a single-order polynomial to each row (wavelength) of the frame. The results from each of the Charon scans were then combined where they overlap. The result is the ratio of the new flat to the old flat with which the Charon data had already been processed by the standard pipeline. After multiplying by the old flat, the new flat was obtained.

To extract the spatially-registered spectra of Arrokoth from the LEISA data frames, a frame-by-frame pointing solution was determined. The changing spacecraft-to-target range during the LEISA scan resulted in a spatial scale that changed from 2.0 to 1.8 km/pixel over the time the scan, with the closer range coinciding with longer wavelengths. The range was taken from predicted SPICE *(91)* kernels for the observation time of each frame, and used to construct a scaled model image of Arrokoth (two sky-plane circles connected by a rectangular neck). The focal plane center of figure, mean albedo, and the full width at half maximum of a Gaussian point spread function was varied until the mean squared difference between a given frame and the model was minimized. Parameter optimization was achieved with simplex minimization. The position angle of the neck and two spheres was assumed to be fixed in all frames. In the case where a pointing solution failed (such as when the target was crossing the edge of the focal plane), the position was estimated via linear interpolation from the nearest successful solutions.

Pixel coordinates in each plane were then converted into a physical frame relative to the center of figure using the range prediction. The [$\Delta x$ (km), $\Delta y$ (km), $\lambda$ (μm)] coordinates of each pixel were stored, where $\Delta x$ and $\Delta y$ are distance from the center of figure along the axes of the detector array, along with the I/F measured by that pixel. A mask of hot pixels and cosmic rays was applied at this stage, and flagged pixels were removed from this data structure. A nearest-neighbor interpolant was constructed to interpolate these I/F measurements over these coordinates. The final spectral cube was constructed by constructing a ($\Delta x$, $\Delta y$, I/F) plane from this interpolant sampled over a spatially-uniform grid sampled at 10 times the average native spatial resolution at each unique wavelength sampled by the measurements.

## Spectral Modeling

Multiple scattering of incident sunlight by a granular surface as a function of composition and texture was simulated using a Hapke model *(31, 92)*. This model was numerically inverted to obtain the best-fitting representations of surface composition and texture, given a set of assumed ingredients and parameters. Because there is an infinite set of potential ingredients and parameters, the resulting best-fitting representations are necessarily not unique solutions, though they can demonstrate that a particular configuration is consistent with the data, and assess spectral differences between regions. Although the MVIC data could provide additional constraints, the models are only applied to LEISA data, because the MVIC and LEISA data sets were obtained at different times and thus at different viewing geometries.

For Hapke parameters, we assumed the single scattering phase function, $P(g)$, from *(3)*, characterized by McGuire & Hapke parameters $b = 0.3178$ and $c = 0.7450$ *(93)*. Hapke's opposition effect and macroscopic roughness parameters were not used (by setting Hapke's $B_0$ and $\bar{\theta}$ parameters to zero). Because all of the spectra being modeled were averages over a range of incidence and emission angles, we computed the models for simplified geometry with incident angle $i = 30°$, emission angle $e = 30°$, and phase angle $g = 10°$. These simplifications are justified by the low signal-to-noise ratio of the data and the coarseness of the spatial registration of the data, owing to the complex shape of the body and changing geometry from spacecraft motion during the LEISA scan. The target was in LEISA's field of view for a little over 4 minutes, during which the spacecraft closed from 32,770 to 29,110 km range from Arrokoth. The changing range is accounted for in our processing of LEISA data, but the phase angle also changed during the scan, and this change is not accounted for. During the scan, the phase angle grew from 12.49° to 12.79°. The start of the scan corresponds to the short wavelength end of the spectral range, so not accounting for this change in geometry could introduce a spurious spectral slope, but with just a 0.3° change over LEISA's spectral range, this is a small effect. This changing geometry over time also results in slightly different geometry between the LEISA data and the LORRI rider obtained during the same scan (see *(3)*, their table S1).

We fitted models to the grand average spectrum incorporating a variety of potential ingredients. We considered the Akaike and Bayesian information criteria (AIC and BIC) to assess whether the inclusion of each particular ingredient was able to reduce $\chi^2$ sufficiently to justify its inclusion. We found only five ingredients to be statistically supported on this basis: tholin, amorphous carbon, and $CH_3OH$ ice, with this model accounting for 92% of the marginalized probability according to the AIC, and to a lesser extent, $NH_3$ ice at 4.5%, and $H_2O$ ice at 3%. According to the BIC, inclusion of $H_2O$ and $NH_3$ are less supported, because it penalizes the inclusion of additional model complexity more harshly than the AIC. We used optical constants for these materials from *(20, 92, 94-96)* for tholin, $CH_3OH$ ice, amorphous carbon, $H_2O$ ice, and $NH_3$ ice, respectively. Tholins from a variety of sources were tried, but according to AIC and BIC criteria, no particular tholin is favored. Titan tholin *(20)* should be seen not as a specific identification but as a generic representative of this broad class of materials. Likewise, amorphous carbon *(94)* is a generic dark opaque material, and is not necessarily an exact match for the dark material present on Arrokoth.

## REX Data Processing

From the shape model of Arrokoth *(3)*, the projected area of Arrokoth as observed from New Horizons at 05:52:14 UTC (time of minimum separation between the high-gain antenna (HGA) boresight and the position of Arrokoth), was 414.4 km$^2$. The distance at this time was 16,700 km. The main beam of the HGA at this distance covered an area about 230 times larger than the target, so the REX measurement was unresolved and only a global determination of the Arrokoth brightness temperature is possible. The observed radio flux is therefore a combination of source plus sky background. The contribution of the background was determined by performing an additional scan along the identical path in right ascension and declination, but one day after the flyby. As viewed from the spacecraft (Fig. 6A), Arrokoth was located on the edge of the Galactic plane in the opposite direction from the Galactic Center, well separated from all bright natural radio sources. With imperfect a priori knowledge of the precise location of Arrokoth, the REX scan was designed to cover all possible time-of-flight errors between ±180 s of the nominal flyby design. This required scanning the HGA over a lengthy arc of 1.2° in RA and 10.2° in Dec, more than adequate to assure covering the slower motion of Arrokoth on the sky (0.5° in RA; 4.2° in Dec). As it turned out, the target was 23 seconds late and thus near the middle of the pre-planned scan range.

The brightness temperature of the sky background within the HGA beam was determined to be only slightly higher than the cosmic microwave background temperature $T_{CMB}$ = 2.73 K. Nevertheless, the radio flux density from the background in each receiver is still about 25 times stronger than the maximum due to the excess thermal emission from Arrokoth. The main contributor to the observation noise, the system temperature of each receiver (REX A~146 K; REX B~137 K), was assumed to remain constant during each scan.

The flux density recorded for the target scan on encounter day is shown together with the background scan recorded one day later (Fig. 6B). The smooth black curve shows the modeled effect of an unresolved Arrokoth with a brightness temperature of 30 K. The target and background scans are plotted over their common declination interval in order to distinguish possible variations in the background. Both curves show only the changes in flux density with respect to the mean value measured during a calibration interval (also indicated in Fig. 6B), defined as those times when the HGA boresight was well separated from Arrokoth. The plotted flux density is a stacked average of the measurements from the two receivers, REX A and REX B, smoothed over a running interval of 0.75° in declination. The smoothing interval was selected as an optimum value: wide enough to reduce the measurement noise, but still narrow enough to capture changes during the roughly 80 seconds when Arrokoth passed through the HGA beam.

Appropriate for the point source case, the observations were fitted to a model with a Gaussian superimposed on a linear background. The best fitting model yielded a change in flux density at maximum of 62 ± 10 Jansky, where the uncertainty accounts for the standard deviation of the entire fit to the data. Converting this to the global brightness temperature of Arrokoth *(36, 37)* yields $T_B$ = 29 ± 5 K, roughly corresponding to a 6-σ detection. Different declination smoothing intervals from 0.25° to 1.25° led to changes in brightness temperature of less than 1 K.

## Thermal models

Our modeling is based on the Arrokoth global shape model *(3)*, reduced to a "low poly" version with $N_f$ = 1962 triangulated facets (Data S1). For a given facet $i$, with surface area $\sigma_i$ and outward pointing normal vector $\mathbf{n}_i$, we determine which other faces $j$ are viewable from it using simple ray-tracing, and from this we develop a logical face network $N_{ij}$, sometimes referred to as the "who-sees-who" list *(38)*. For each $i$-$j$ pair, we assess their relative distance $r_{ij}$ with corresponding unit vector $\mathbf{r}_{ij}$. This procedure is nominally an $O(N_f^2)$ calculation, but is done only once for any given shape model and is subsequently used for all derived thermal models. This methodology accounts for large scale roughness expressed in the faceted shape model, but does not account for smaller scale roughness which can also be expected to affect surface temperatures.

We adopt a surface plus one-dimensional subsurface model to describe the near surface energy balance in which losses to radiation and subsurface conduction is balanced by the net received energy, $\Phi_i$, which is the radiative losses plus the gains from both solar insolation and surface re-radiation,

$$k\partial_z T_i|_{z=0} + \epsilon_i \sigma_B T_i^4 \;=\; \Phi_i \;=\; (1-A_i)F_i + \sum_j \epsilon_j \sigma_B K_{ij} T_j^4,$$
$$F_i \;=\; F_*(\mathbf{n}_* \cdot \mathbf{n}_i),$$

(S1)

where $z$ is depth and the summation is over all non-zero $N_{ij}$. $T_i$ is the surface temperature of element $i$. The local emissivity is $\epsilon_i$ (assumed to be 0.9, following *(40, 43)*), $F_i$ is the non-self-blocked insolation received by surface element *i* from the Sun located in the sky by unit vector **n**$_*$, (1-$A_i$)$F_i$ is the absorbed solar radiation flux (taking into account self-shadowing), and $A_i$ the bond albedo, measured to be 0.06 *(1, 3)*. The scaled radiated power emanating from facet *j* and received by facet *i* is quantified by the following expression:

$$K_{ij} \;=\; \sigma_i N_{ij} \frac{(\mathbf{n}_i \cdot \mathbf{r}_{ij})(-\mathbf{n}_j \cdot \mathbf{r}_{ij})}{2\pi r_{ij}^2}. \tag{S2}$$

We adopt the 1D heat equation for the subsurface,

$$\rho C_p \partial_t \theta_i \;=\; k \partial_z^2 \theta_i, \tag{S3}$$

where $\rho$ is the material density, $C_p$ is its heat capacity, $\theta_i$ is the depth dependent (*z*) temperature of facet *i* where surface temperature is denoted by $T_i = \theta_i$ (at *z* = 0), and finally *k* is the material conductivity. Solutions to Eq. S3 are developed in the time-asymptotic limit whereby we impose the condition that the conductive flux goes to zero sufficiently deep underneath the surface *(39)*. This amounts to representing the solution for the temperature for z < 0 as a truncated sum in powers of seasonal frequency, i.e.,

$$\theta_i \;=\; \bar{T}_i + \sum_{n=1,N} e^{\kappa_n z}\left(A_n e^{i\omega_0 m t} + \text{c.c.}\right);$$

$$\kappa_n \;\equiv\; (1+i)\sqrt{\frac{n\omega_0 \rho C_p}{2k}}; \tag{S4}$$

where $\bar{T}_i$ is the local mean temperature of element *i*. This solution approach has also been used in the thermal response analysis of Martian sand dunes *(97)*. We construct solutions based on daily averaged solar insolation and proceed by building a received insolation profile for 298 evenly selected times over the course of ~298 year orbit of Arrokoth. We truncate the number of frequencies retained in Eq. S4 to about 75 so as to minimize the possibility of incurring de-aliasing errors in the solution procedure. We then build solutions to Eqs. S1 and S3 via an iterative procedure. We start by solving Eq. S1 for $T_i$ with *k* set to zero and calculate a first iteration of $\Phi_i$. This solution involves the inversion of a sparse matrix. This can be done via a forward iterative Gauss procedure because the matrix is well-conditioned. Then we correct $T_i$ by solving Eq. S3 using this first guess for $\Phi_i$, as the flux boundary condition $k\partial_z\theta_i|_{z=0} + \epsilon_i\sigma_B\theta_i^4|_{z=0} = \Phi_i$. This provides a corrected surface temperature $T_i = \theta_i|_{z=0}$ which is input back into Eq. S1 whereupon an updated value of $\Phi_i$ is assessed and used as an updated upper boundary condition used to solve Eq. S3. This iterates until a converged solution is reached which is usually assessed to be when the maximum change between successive iterations is sufficiently small, i.e., when max($\Delta\Phi_i/\Phi_i$) < $10^{-4}$, where $\Delta\Phi_i$ is the difference between successive iterations of $\Phi_i$.

Following *(1, 40)*, we assumed thermal inertia $\Gamma$ = 2.5 J m$^{-2}$ s$^{-1/2}$ K$^{-1}$. The resulting 16 K model mean surface temperature across the face of Arrokoth oriented toward New Horizons during the REX CA08 observation is far below the 29 ± 5 K brightness temperature recorded by REX. However, thermal radiation tends to emerge from a range of depths within a surface, potentially extending to many times the wavelength below the surface. Additionally, we do not know the X-band emissivity ($\epsilon$) of Arrokoth's surface that provides the link between kinetic and brightness temperatures. To explore the range of parameters permitted by the REX observation, we re-ran our model with $\Gamma$ ranging from 0.5 to 250 J m$^{-2}$ s$^{-1/2}$ K$^{-1}$ and computed temperatures at depths from the surface down to 4 m below the surface (~100 wavelengths), as shown in Fig. S1.

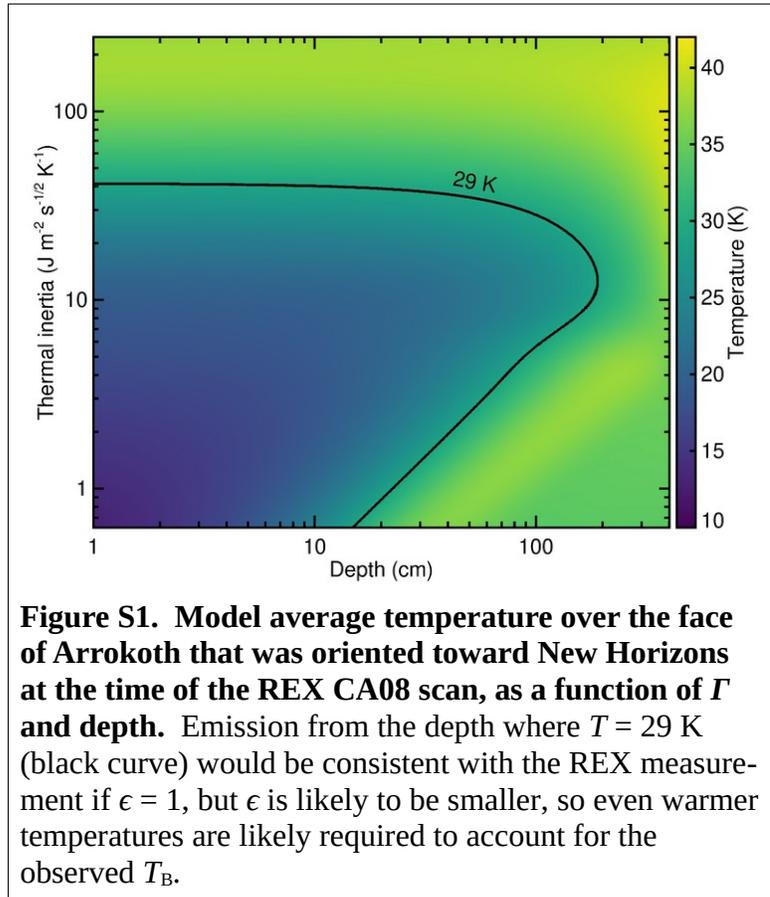

The observed REX thermal emission is consistent with low thermal inertia and emission emerging from tens of cm below the surface, but is also consistent with higher thermal inertia and emission from closer to the surface. For instance, at the nominal 2.5 J m$^{-2}$ s$^{-1/2}$ K$^{-1}$ thermal inertia, the mean temperature over the observed hemisphere is 29 K at a mean depth of ~50 cm (12 REX wavelengths). For less than unit emissivity, warmer temperatures are needed to account for the observed $T_B$, consistent with emission from greater depths. With $\epsilon = 0.8$, the corresponding kinetic temperature would be 36 K, which is reached at a depth of ~1 m. The maximum temperatures seen for any thermal inertia and depth combination are in the ~41 K range, implying a lower limit of 0.7 for the REX wavelength emissivity.

**Figure S1. Model average temperature over the face of Arrokoth that was oriented toward New Horizons at the time of the REX CA08 scan, as a function of $\Gamma$ and depth.** Emission from the depth where $T = 29$ K (black curve) would be consistent with the REX measurement if $\epsilon = 1$, but $\epsilon$ is likely to be smaller, so even warmer temperatures are likely required to account for the observed $T_B$.

For the higher thermal inertia solutions, the required value of $\Gamma$ would be >40 J m$^{-2}$ s$^{-1/2}$ K$^{-1}$ to reconcile the surface temperature with the REX observation with unit emissivity at REX wavelengths. Surface and subsurface temperatures are similar for the high thermal inertia models, so the higher thermal inertia models place less constraint on the depth sampled by the 4.2 cm the emission. A high thermal inertia is inconsistent with the statistical estimates from *(40)*, but those estimates are for diurnal timescales rather than seasonal timescales. Higher thermal conduction at depth could enable seasonal thermal inertias to be much higher than the values relevant for diurnal timescales. Additionally, the *(40)* sample does not include small CCKBOs in Arrokoth's size range, so it might not even be relevant to such objects.

Effects of unresolved surface roughness can be important in some thermal environments. These were explored by *(98)* through use of effective values of the albedo and emissivity.

Arrokoth's bond albedo has been directly measured as 0.06 *(1, 3)*, so we leave it unchanged, but we tried higher values of emissivity $\epsilon = 0.95$ and $\epsilon = 1.0$ consistent with a rough surface. These led to slightly lower temperatures by up to 1 K across the observed face of Arrokoth, not enough to materially affect our conclusions.

**Data S1 (separate file).**

A low facet count version of the Arrokoth shape model *(3)* used in our thermal models is provided as an ASCII file consisting of 1,039 lines representing vertices starting with the character v, followed by 1,962 lines representing facets starting with the character f. Each vertex line has three white-space separated floating point values representing *x, y, z* coordinates in units of km. The axes are aligned with the shape's principal axes of inertia and the origin corresponds to the center of mass assuming a uniform internal density. Each facet line has three white-space separated integers specifying three vertices to define the corners of that triangular facet, in order such that the right hand rule points in the direction of the exterior of the shape. The index of the first vertex line is 1 and the last is 1,039.